\documentclass[journal]{IEEEtran}

\usepackage[english]{babel}

\usepackage{ifpdf}

\usepackage{cite} 
\usepackage{url}
\usepackage{hyperref}

\ifCLASSINFOpdf
	\usepackage[pdftex]{graphicx}
	\graphicspath{{./figures/}}
\else
	\usepackage[dvips]{graphicx}
	\graphicspath{./figures/}
\fi
\usepackage{color}
\usepackage{pgf, tikz, pgfplots}
\usetikzlibrary{shapes, arrows, automata}
\usetikzlibrary{calc,hobby,decorations}

\usepackage[cmex10]{amsmath}
\usepackage{amsfonts, amssymb, amsthm}
\usepackage{mathrsfs}

\usepackage{algorithm,algpseudocode}
	\algnewcommand{\LeftComment}[1]{\Statex \(\triangleright\) #1}

\usepackage{enumerate}
\usepackage{multirow}
\usepackage{rotating}
\usepackage{subcaption}
\usepackage{needspace}

\newcommand{\myparagraph}[1]{\needspace{1\baselineskip}\medskip\noindent {\bf #1}}

\input{mySymbol.sty}

\definecolor{penndarkestblue}{cmyk}{1,0.74,0,0.77}
\definecolor{penndarkerblue}{cmyk}{1,0.74,0,0.70}
\definecolor{pennblue}{cmyk}{0.99,0.66,0,0.57} 
\definecolor{pennlighterblue}{cmyk}{0.98,0.44,0,0.35}
\definecolor{pennlightestblue}{cmyk}{0.38,0.17,0,0.17} 

\definecolor{penndarkestred}{cmyk}{0,1,0.89,0.66}
\definecolor{penndarkerred}{cmyk}{0,1,0.88,0.55}
\definecolor{pennred}{cmyk}{0,1,0.83,0.42} 
\definecolor{pennlighterred}{cmyk}{0,1,0.6,0.24}
\definecolor{pennlightestred}{cmyk}{0,0.43,0.26,0.12} 

\definecolor{penndarkestgreen}{cmyk}{1,0,1,0.68}
\definecolor{penndarkergreen}{cmyk}{1,0,1,0.57}
\definecolor{penngreen}{cmyk}{1,0,1,0.44} 
\definecolor{pennlightergreen}{cmyk}{1,0,1,0.25}
\definecolor{pennlightestgreen}{cmyk}{0.43,0,0.43,0.13}

\definecolor{penndarkestorange}{cmyk}{0,0.65,1,0.49}
\definecolor{penndarkerorange}{cmyk}{0,0.65,1,0.33}
\definecolor{pennorange}{cmyk}{0,0.54,1,0.24} 
\definecolor{pennlighterorange}{cmyk}{0,0.32,1,0.13}
\definecolor{pennlightestorange}{cmyk}{0,0.15,0.46,0.06}
	
\definecolor{penndarkestpurple}{cmyk}{0,1,0.11,0.86}
\definecolor{penndarkerpurple}{cmyk}{0,1,0.13,0.82}
\definecolor{pennpurple}{cmyk}{0,1,0.11,0.71} 
\definecolor{pennlighterpurple}{cmyk}{0,1,0.05,0.46}
\definecolor{pennlightestpurple}{cmyk}{0,0.35,0.02,0.23}
	
\definecolor{pennyellow}{cmyk}{0,0.20,1,0.05} 
\definecolor{pennlightgray1}{cmyk}{0,0,0,0.05}
\definecolor{pennlightgray3}{cmyk}{0.01,0.01,0,0.18}
\definecolor{pennmediumgray1}{cmyk}{0.04,0.03,0,0.31}
\definecolor{pennmediumgray4}{cmyk}{0.08,0.06,0,0.54}
\definecolor{penndarkgray2}{cmyk}{0.09,0.07,0,0.71}
\definecolor{penndarkgray4}{cmyk}{0.1,0.1,0,0.92}

\def\Tr{\mathsf{T}}
\def\Hr{\mathsf{H}}

\def\dc{\text{dc}}

\newtheorem{remark}{\hspace{0pt}\bf Remark}

\begin{document}

\title{Convolutional Neural Network Architectures for Signals Supported on Graphs}

\author{Fernando~Gama,~
        Antonio~G.~Marques,~
        Geert~Leus,~
        and~Alejandro~Ribeiro
\thanks{Supported by NSF CCF 1717120, ARO W911NF1710438, ARL DCIST CRA W911NF-17-2-0181, ISTC-WAS and Intel DevCloud; and Spain MINECO grants No TEC2013-41604-R and TEC2016-75361-R. F. Gama and A. Ribeiro are with the Dept. of Electrical and Systems Eng., Univ. of Pennsylvania., A. G. Marques is with the Dept. of Signal Theory and Comms., King Juan Carlos Univ., G. Leus is with the Dept. of Microelectronics, Delft Univ. of Technology.  Email: \{fgama,aribeiro\}@seas.upenn.edu, antonio.garcia.marques@urjc.es, and g.j.t.leus@tudelft.nl.
}
}

\markboth{IEEE TRANSACTIONS ON SIGNAL PROCESSING (ACCEPTED)}%
{Convolutional Neural Networks Architectures for Signals Supported on Graphs}

\maketitle

\begin{abstract}
Two architectures that generalize convolutional neural networks (CNNs) for the processing of signals supported on graphs are introduced. We start with the selection graph neural network (GNN), which replaces linear time invariant filters with linear shift invariant graph filters to generate convolutional features and reinterprets pooling as a possibly nonlinear subsampling stage where nearby nodes pool their information in a set of preselected sample nodes. A key component of the architecture is to remember the position of sampled nodes to permit computation of convolutional features at deeper layers. The second architecture, dubbed aggregation GNN, diffuses the signal through the graph and stores the sequence of diffused components observed by a designated node. This procedure effectively aggregates all components into a stream of information having temporal structure to which the convolution and pooling stages of regular CNNs can be applied. A multinode version of  aggregation GNNs is further introduced for operation in large scale graphs. An important property of selection and aggregation GNNs is that they reduce to conventional CNNs when particularized to time signals reinterpreted as graph signals in a circulant graph. Comparative numerical analyses are performed in a source localization application over synthetic and real-world networks. Performance is also evaluated for an authorship attribution problem and text category classification. Multinode aggregation GNNs are consistently the best performing GNN architecture.
\end{abstract}


\IEEEpeerreviewmaketitle


\section{Introduction} \label{sec_intro}



We consider signals with irregular structure and describe their underlying topology with a graph whose edge weights capture a notion of expected similarity or proximity between signal components expressed at nodes \cite{sandryhaila13-dspg, sandryhaila14-freq, shuman13-mag, sandryhaila14-mag}. Of particular importance in this paper is the interpretation of matrix representations of the graph as shift operators that can be applied to graph signals. Shift operators represent local (one-hop neighborhood) operations on the graph, and allow for different generalizations of convolution, sampling and reconstruction. These generalizations stem either from representations of graph filters as polynomials in the shift operator \cite{sandryhaila13-dspg, segarra17-linear, shuman18-chebyshev} or from the aggregation of sequences generated through successive application of the shift operator \cite{marques16-aggregation}. They not only capture the intuitive idea of convolution, sampling and reconstruction as local operations but also share some other interesting theoretical properties \cite{segarra17-linear, sandryhaila13-dspg, sandryhaila14-freq}. Our goal here is to build on these definitions to generalize Convolutional (C) neural networks (NNs) to graph signals.

\begin{figure*}[t]
\centering
	\begin{subfigure}{.3\textwidth}
		\centering
		\includegraphics[width=\textwidth]{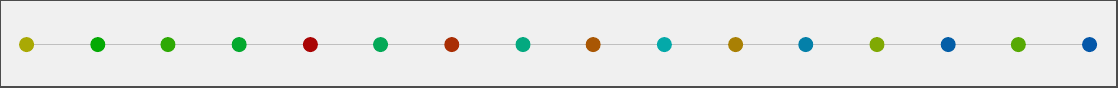}
		\caption{input}
		\label{layer1input}
	\end{subfigure}
	\hfill
	\begin{subfigure}{.3\textwidth}
		\centering
		\includegraphics[width=\textwidth]{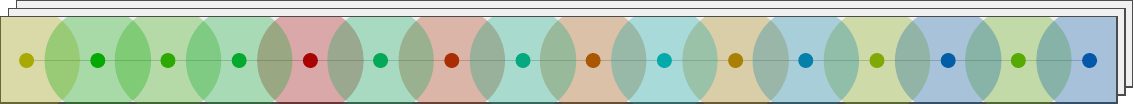} 
		\caption{convolution}
		\label{layer1conv}
	\end{subfigure}
	\hfill
	\begin{subfigure}{.3\textwidth}
		\centering
		\includegraphics[width=\textwidth]{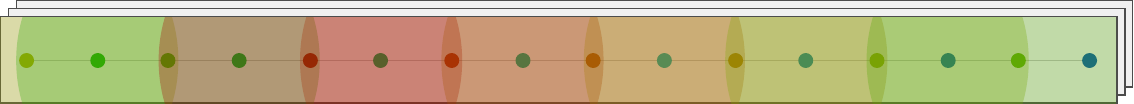} 
		\caption{pooling}
		\label{layer1pool}
	\end{subfigure}
	\\ \vspace{0.5cm}
	\begin{subfigure}{.16\textwidth}
		\centering
		\includegraphics[width=\textwidth]{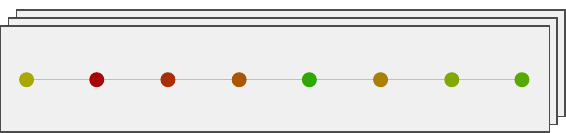}
		\caption{input}
		\label{layer2input}
	\end{subfigure}
	\hfill
	\begin{subfigure}{.16\textwidth}
		\centering
		\includegraphics[width=\textwidth]{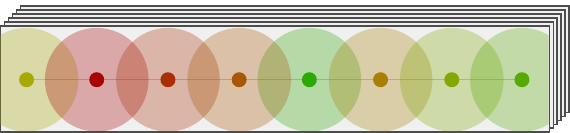}
		\caption{convolution}
		\label{layer2conv}
	\end{subfigure}
	\hfill
	\begin{subfigure}{.16\textwidth}
		\centering
		\includegraphics[width=\textwidth]{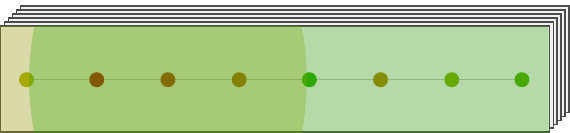}
		\caption{pooling}
		\label{layer2pool}
	\end{subfigure}
	\hfill
	\begin{subfigure}{.05\textwidth}
		\centering
		\includegraphics[width=\textwidth]{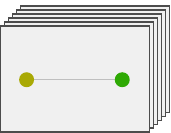}
		\caption{}
		\label{layer3input}
	\end{subfigure}
	\hfill
	\begin{subfigure}{.05\textwidth}
		\centering
		\includegraphics[width=\textwidth]{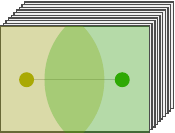}  
		\caption{}
		\label{layer3conv}
	\end{subfigure}
	\hfill
	\begin{subfigure}{.05\textwidth}
		\centering
		\includegraphics[width=\textwidth]{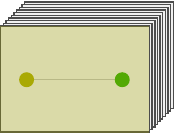}
		\caption{}
		\label{layer3pool}
	\end{subfigure}
\caption{Convolutional Neural Networks. \subref{layer1input} Consider the input to be a discrete time signal, represented by a succession of signal values. \subref{layer1conv} Convolve this signal with a filter to obtain corresponding features [cf. \eqref{eqn:conv_time}]. The color disks centered at each node symbolize the convolution operation. \subref{layer1pool} Apply pooling [cf. \eqref{eqn_group_nonlinearity}]. The color disks symbolize the reach of the pooling operation (the number of samples that are pooled together) \subref{layer2input} Downsample to obtain a discrete time signal of smaller size [cf. \eqref{eqn_downsampling}]. \subref{layer2conv}-\subref{layer3pool} Repeat the application of convolution and pooling, trading off the temporal dimension for more features.}
\label{fig:regular_cnn}
\end{figure*}

CNNs consist of layers that are sequentially composed, each of which is itself the composition of convolution and pooling operations (Section \ref{sec:regular} and Figure~\ref{fig:regular_cnn}). The input to a layer is a multichannel signal composed of features extracted from the previous layer, or the input signal itself at the first layer. The main step in the convolution stage is the processing of each feature with a bank of linear time invariant filters (Section \ref{sec_cnn_convolution}). To keep complexity under control and avoid the number of intermediate features growing exponentially, the outputs of some filters are merged via simple pointwise summations. In the pooling stage we begin by computing local summaries in which feature components are replaced with a summary of their values at nearby points (Sec. \ref{sec_cnn_pooling}). These summaries can be linear, e.g., a weighted average of adjacent components, or nonlinear, e.g., the maximum value among adjacent components. Pooling also involves a subsampling of the summarized outputs. This subsampling reduces dimensionality with a (small) loss of information because the summarizing function is a low-pass operation. The output of the layer is finally obtained by application of a pointwise nonlinear activation function to produce features that become an input to the next layer. This is an architecture that is both simple to implement \cite{najafabadi15-cnnbigdata}, and simple to train \cite{rumelhart86-backprop}. Most importantly, their performance in regression and classification is remarkable to the extent that CNNs have become the standard tool in machine learning to handle such inference tasks \cite{lecun15-deeplearning, lecun10-vision, greenspan16-medical}.

As it follows from the above description, a CNN layer involves five operations: (i) Convolution with linear time invariant filters. (ii) Summation of different features. (iii) Computation of local summaries. (iv) Subsampling. (v) Activation with a pointwise nonlinearity. A graph (G)NN is an architecture adapted to graph signals that generalizes these five operations. Operations (ii) and (v) are pointwise, therefore independent of the underlying topology, so that they can be applied without modification to graph signals. Generalizing (iii) is ready because the notion of adjacent components is well defined by graph neighborhoods. Generalization of operation (i) is not difficult in the context of graph signal processing advances whereby linear time invariant filters are particular cases of linear shift invariant graph filters. This has motivated the definition of graph (G) NNs with convolutional features computed from shift invariant graph filters, an idea that was first introduced in \cite{bruna14-deepspectralnetworks} and further explored in \cite{henaff15-deepgraph, atwood16-diffusion, defferrard17-cnngraphs, du17-topoadapt, kipf17-classifgcnn, gama18-mimo}. Architectures based on receptive fields, which are different but conceptually similar to graph filters, have also been proposed \cite{niepert16-learningcnn, pasdeloup17-approxtrans, velickovic18-graphattentionnetworks}. However, generalization of operation (iv) has proven more challenging because once the signal is downsampled, it is not easy to identify a coarsened graph to connect the components of the subsampled signal. The use of multiscale hierarchical clustering has been proposed to produce a collection of smaller graphs \cite{bruna14-deepspectralnetworks, henaff15-deepgraph, defferrard17-cnngraphs} but it is not clear which clustering or coarsening criteria is appropriate for GNN architectures. The difficulty of designing and implementing proper pooling is highlighted by the fact that several works exclude the  pooling stage altogether \cite{du17-topoadapt, niepert16-learningcnn, pasdeloup17-approxtrans, gama18-nvgf}. 

In this paper we propose two different GNN architectures, selection GNNs and aggregation GNNs, that include convolutional and pooling stages but bypass the need to create a coarsened graph. In selection GNNs (Sec. \ref{sec:selection} and Fig. \ref{fig:selection_cnn}) we replace convolutions with linear shift invariant filters and replace regular sampling with graph selection sampling. In the first layer of the selection GNN, linear shift invariant filters are well defined as polynomials on the given graph. At the first pooling stage, however, we sample a smaller number of signal components and face the challenge of computing a graph to describe the topology of the subsampled signal. Our proposed strategy is to bypass the computation of a coarsened graph by using zero padding (Sec. \ref{sec_selection_convolution}). This simple technique permits computation of features that are convolutional on the input graph. The pooling stage is modified to aggregate information in multihop neighborhoods as determined by the structure of the original graph and the sparsity of the subsampled signal (Sec. \ref{sec_selection_pooling}).

In aggregation GNNs we borrow ideas from aggregation sampling \cite{marques16-aggregation} to create a signal with temporal structure that incorporates the topology of the graph (Sec. \ref{sec:aggregation} and Fig. \ref{fig_aggregation}). This can be accomplished by focusing on a designated node and considering the local sequence that is generated by subsequent applications of the graph shift operator. This is a signal with a temporal structure because it reflects the propagation of a diffusion process. Yet, it also captures the topology of the graph because subsequent components correspond to the aggregation of information in nested neighborhoods of increasing reach. Aggregation GNNs apply a regular CNN to the diffusion signal observed at the designated node. 

We finally introduce a multinode version of aggregation GNNs, where several regular CNNs are run at several designated nodes (Sec. \ref{sec_aggregation_multinode} and Fig. \ref{fig_multinode}). The resulting CNN outputs are diffused in the input graph to generate another sequence with temporal structure at a smaller subset of nodes to which regular CNNs are applied in turn. We can think of multinode aggregation GNNs as composed of inner and outer layers. Inner layers are regular CNN layers. Output layers stack CNNs joined together by a linear diffusion process. Multinode aggregation GNNs are consistently the best performing GNN architecture (Sec. \ref{sec:sims}). We remark that aggregation GNNs, as well as selection GNNs are proper generalizations of conventional CNNs because they both reduce to a CNN architecture when particularized to a cyclic graph.

The proposed architectures are applied to the problems of localizing the source of a diffusion process on synthetic networks (Sec.~\ref{subsec_sourceloc}) as well as on real-world social networks (Sec.~\ref{subsec_fb}). Performance is additionally evaluated on problems of authorship attribution (Sec.~\ref{subsec_author}) and classification of articles of the \texttt{20NEWS} dataset (Sec.~\ref{subsec_20news}), involving real datasets. Results are compared to those obtained from a graph coarsening architecture using a multiscale hierarchical clustering scheme \cite{defferrard17-cnngraphs}. The results are encouraging and show that the multinode approach consistently outperforms the other architectures.

\noindent \emph{Notation:} The $n$-th component of a vector $\bbx$ is denoted as $[\bbx]_{n}$. The $(m,n)$ entry of a matrix $\bbX$ is $[\bbX]_{mn}$. The vector $\bbx:=[\bbx_1; \ldots; \bbx_n]$ is a column vector stacking the column vectors $\bbx_n$. When $\bbn$ denotes a set of subindices, $|\bbn|$ is the number of elements in $\bbn$ and  $[\bbx]_{\bbn}$ denotes the column vector formed by the elements of $\bbx$ whose subindices are in $\bbn$. The vector $\bbone$ is the all-ones vector.


\section{Convolutional Neural Networks} \label{sec:regular}


%
Given a training set $\ccalT := \{(\bbx,\bby)\}$ formed by inputs $\bbx$ and their associated outputs $\bby$, a learning algorithm produces a representation (mapping) that can estimate the output $\hby$ that should be assigned to an input $\hbx\notin\ccalT$. NNs produce a representation using a stacked layered architecture in which each layer composes a linear transformation with a pointwise nonlinearity \cite{goodfellow16-deeplearn}. Formally, the first layer of the architecture begins with a linear transformation to produce the intermediate output $\bbu_1 := \bbA_{1}\bbx_{0} = \bbA_{1}\hbx$ followed by a pointwise nonlinearity to produce the first layer output $\bbx_1 := \sigma_1(\bbu_1) = \sigma_1(\bbA_{1}\bbx_0)$. This procedure is applied recursively so that at the $\ell$th layer we compute the transformation
\begin{equation}\label{eqn_nn_layers}
    \bbx_\ell := \sigma_\ell(\bbu_\ell) := \sigma_\ell(\bbA_{\ell}\bbx_{\ell-1}).
\end{equation}
In an architecture with $L$ layers, the input $\hbx=\bbx_0$ is fed to the first layer and the output $\hby = \bbx_L$ is read from the last layer \cite{kuo17-recos}. Elements of the training set $\ccalT$ are used to find matrices $\bbA_{\ell}$ that optimize a training cost of the form $\sum_{(\bbx,\bby)\in\ccalT} f(\bby, \bbx_L)$, where $f(\bby, \bbx_L)$ is a fitting metric that assess the difference between the NN's output $\bbx_L$ produced by input $\bbx$ and the desired output $\bby$ stored in the training set. Computation of the optimal NN coefficients $\bbA_\ell$ is typically carried out by stochastic gradient descent, which can be efficiently computed using the backpropagation algorithm \cite{rumelhart86-backprop}.

The NN architecture in \eqref{eqn_nn_layers} is a multilayer perceptron composed of fully connected layers \cite{kuo17-recos}. If we denote as $M_\ell$ the number of entries of the output of layer $\ell$, the matrix $\bbA_{\ell}$ contains $M_{\ell}\times M_{\ell-1}$ components. This, likely extremely, large number of parameters not only makes training challenging but empirical evidence suggests that it leads to overfitting \cite{huang17-densecnn}. CNNs resolve this problem with the introduction of two operations: Convolution and pooling.

%

%
\subsection{Convolutional Features}\label{sec_cnn_convolution}

To describe the creation of convolutional features write the output of the $(\ell-1)$st layer as $\bbx_{\ell-1} := [\bbx^{1}_{\ell-1}; \ldots; \bbx^{F_{\ell-1}}_{\ell-1}]$. This decomposes the $M_{\ell-1}$-dimensional output of the $(\ell-1)$st layer as a stacking of $F_{\ell-1}$ features of dimension $N_{\ell-1}$. This collection of features is the input to the $\ell$th layer. Likewise, the intermediate output $\bbu_{\ell}$ can be written as a collection of $F_{\ell}$ features $\bbu_\ell := [\bbu^{1}_{\ell}; \ldots; \bbu^{F_\ell}_{\ell}] $ where $\bbu^{f}_\ell$ is of length $N_{\ell-1}$ and is obtained through convolution and linear aggregation of features $\bbx^{g}_{\ell-1}$ of the previous layer, $g=1,\ldots,F_{\ell-1}$. Specifically, let $\bbh_{\ell}^{fg} := [\ [\bbh_{\ell}^{fg}]_{0}; \ldots; [\bbh_{\ell}^{fg}]_{K_{\ell}-1} \ ]$ be the coefficients of a $K_\ell$-tap linear time invariant filter that is used to process the $g$th feature of the $(\ell-1)$st layer to produce the intermediate feature $\bbu_{\ell}^{fg}$ at layer $\ell$. Since the filter is defined by a convolution, the components of $\bbu_{\ell}^{fg}$ are explicitly given by
\begin{align}\label{eqn:conv_time}
   \Big[\bbu_{\ell}^{fg}\Big]_n 
       :=   \Big[\bbh_{\ell}^{fg} \ast \bbx_{\ell-1}^{g}\Big]_{n}  
	    =   \sum_{k=0}^{K_{\ell}-1}  \Big[ \bbh_{\ell}^{fg}  \Big]_{k}  \,
	        \Big[ \bbx_{\ell-1}^{g} \Big]_{n-k} ,
\end{align}
where we consider that: i) the output has the same size than the input and ii) the convolution \eqref{eqn:conv_time} is circular to account for border effects. After evaluating the convolutions in \eqref{eqn:conv_time}, the $\ell$th layer features $\bbu_{\ell}^{f}$ are computed by aggregating the intermediate features $\bbu_{\ell}^{fg}$ associated with each of the previous layer features $\bbx_{\ell-1}^{g}$ using a simple summation, 
\begin{equation} \label{eqn:agg_features}
   \bbu_{\ell}^{f}
       \ :=\ \sum_{g=1}^{F_{\ell-1}} \bbu_{\ell}^{fg}
	   \  =\ \sum_{g=1}^{F_{\ell-1}} \bbh_{\ell}^{fg} \ast \bbx_{\ell-1}^{g} .
\end{equation}
The vector $\bbu_\ell := [\bbu^{1}_{\ell}; \ldots; \bbu^{F_\ell}_{\ell}]$ obtained from \eqref{eqn:conv_time} and \eqref{eqn:agg_features} represents the output of the linear operation of the $\ell$th layer of the CNN [cf. \eqref{eqn_nn_layers}]. Although not explicitly required, the number of features $F_{\ell}$ and the number of filter taps $K_{\ell}$ are typically much smaller than the dimensionality $M_{\ell-1}$ of the features $\bbx_{\ell-1}$ that are processed by the $\ell$th layer. This reduces the number of learnable parameters from $M_{\ell}\times M_{\ell-1}$ in \eqref{eqn_nn_layers} to $K_{\ell}\times F_{\ell}\times F_{\ell-1}$ simplifying training and reducing overfitting. 

%
\subsection{Pooling}\label{sec_cnn_pooling}

The features $\bbu_{\ell}^{fg}$ in \eqref{eqn:conv_time} and their consolidated counterparts $\bbu_{\ell}^{f}$ in \eqref{eqn:agg_features} have $N_{\ell-1}$ components. This number of components is reduced to $N_{\ell}$ at the pooling stage in which the values of a group of neighboring elements are aggregated to a single scalar using a possibly nonlinear summarization function $\rho_\ell$. To codify the locality of $\rho_\ell$, we define, with a slight abuse of notation, $\bbn_\ell$ as a vector containing the indexes associated with index $n$ -- e.g., use $\bbn_\ell=[n-1; n; n+1]$ to group adjacent components -- and define the signal $\bbv^{f}_{\ell}$ with components
\begin{equation}\label{eqn_group_nonlinearity}
   \Big[\bbv^{f}_{\ell}\Big]_n 
            = \rho_{\ell}\left( \Big[\bbu_{\ell}^f\Big]_{\bbn_\ell}\right).
\end{equation}
The summarization function $\rho_{\ell}$ in \eqref{eqn_group_nonlinearity} acts as a low-pass operation and the most common choices are the maximum $\rho_{\ell}( [\bbu_{\ell}^f]_{\bbn_\ell}) = \max( [\bbu_{\ell}^f]_{\bbn_\ell})$ and the average  $\rho_{\ell}( [\bbu_{\ell}^f]_{\bbn_\ell}) = \bbone^{\Tr} [\bbu_{\ell}^f]_{\bbn_\ell}/ |\bbn_\ell|$ \cite{wiatowski17-maththeory}.

To complete the pooling stage we follow $\eqref{eqn_group_nonlinearity}$ with a downsampling operation. For that matter, we define the sampling matrix $\bbC_{\ell}$ as a fat binary matrix with $N_{\ell-1}$ columns and $N_{\ell}$ rows, which are selected from the rows of the identity matrix. When the sampling matrix $\bbC_{\ell}$ is \textit{regular}, the nonzero entries follow the pattern $[\bbC_{\ell}]_{mn}=1$ if $n$ can be written as $n=(N_{\ell-1}/N_{\ell})m$ and zero otherwise; hence, the product $\bbC_{\ell}\bbv^{f}_{\ell}$ selects one out of every $(N_{\ell-1}/N_{\ell})$ components of $\bbv^{f}_{\ell}$. Downsampling is composed with a pointwise nonlinearity to produce the $\ell$th layer features
\begin{equation}\label{eqn_downsampling}
   \bbx^{f}_{\ell} =  \sigma_\ell \left(\bbC_{\ell}\bbv^{f}_{\ell} \right).
\end{equation}
The compression or downsampling factor $(N_{\ell-1}/N_{\ell})$ is often matched to the local summarization function $\rho_{\ell}$ so that the set $\bbn_\ell$ contains $(N_{\ell-1}/N_{\ell})$ adjacent indexes. We further note that although we defined \eqref{eqn_group_nonlinearity} for all $n$, in practice, we only compute the components of $\bbv^{f}_{\ell}$ that are to be selected by the sampling matrix $\bbC_\ell$. In fact, it is customary to combine \eqref{eqn_group_nonlinearity} and \eqref{eqn_downsampling} to simply write $[\bbx^{f}_{\ell}]_n = \sigma_l (\rho_{\ell}( [\bbu_{\ell}^f]_{\bbn_\ell})$ for $n$ in the selection set. Separating the nonlinearity in \eqref{eqn_group_nonlinearity} from the downsampling operation in \eqref{eqn_downsampling} is convenient to elucidate pooling strategies for signals on graphs.

Equations \eqref{eqn:conv_time}-\eqref{eqn_downsampling} complete the specification of the CNN architecture. We begin at each layer with the input $\bbx_{\ell-1} := [\bbx^{1}_{\ell-1}; \ldots; \bbx^{F_{\ell-1}}_{\ell-1}]$. Features are fed to parallel convolutional channels to produce the features  $\bbu_{\ell}^{fg}$ in \eqref{eqn:conv_time} and consolidated into the features $\bbu_{\ell}^{f}$ in \eqref{eqn:agg_features}. These features are fed to the local summarization function $\rho_{\ell}$ to produce features $\bbv^{f}_{\ell}$ [cf. \eqref{eqn_group_nonlinearity}] which are then downsampled and processed by the pointwise activation nonlinearity $\sigma_{\ell}$ to produce the features $\bbx^{f}_{\ell}$ [cf. \eqref{eqn_downsampling}]. The output of the $\ell$th layer is the vector $\bbx_\ell := [\bbx^{1}_{\ell}; \ldots; \bbx^{F_\ell}_{\ell}]$ that groups the features in \eqref{eqn_downsampling}. We point out for completeness that the $L$th layer is often a fully connected layer in the mold of \eqref{eqn_nn_layers} that does not abide to the convolutional and pooling paradigm of \eqref{eqn:conv_time}-\eqref{eqn_downsampling}. Thus, the $L$th layer produces an arbitrary (non convolutional) linear combination of $F_{L-1}$ features to produce the final $F_{L}$ scalar features $\bbx_L$. The output of this readout layer provides the estimate $\hby = \bbx_L$ that is associated with the input $\hbx=\bbx_0$ fed to the first layer.

%
\subsection{Signals on Graphs}\label{sec_cnn_graph signals}

There is overwhelming empirical evidence that CNNs are superb representations of signals defined in regular domains such as time series and images \cite{lecun15-deeplearning}. Our goal in this paper is to contribute to the extension of these architectures to signals supported in irregular domains described by arbitrary graphs. Consider then a weighted graph with $N$ nodes, edge set $\ccalE$ and weight function $\ccalW: \ccalE \to \reals$. We endow the graph with a shift operator $\bbS$, which is an $N\times N$ square matrix having the same sparsity pattern of the graph; i.e., we can have $[\bbS]_{mn}\neq 0$ if and only if $(n,m) \in \ccalE$ or $m=n$. The shift operator is a stand in for one of the matrix representations of the graph. Commonly used shift operators include the adjacency matrix $\bbA$ with nonzero elements $[\bbA]_{mn}=\ccalW(n,m)$ for all $(n,m)\in \ccalE$, the Laplacian $\bbL:=\diag(\bbA \bbone)-\bbA$ and their normalized counterparts $\barbA$ and $\barbL$ \cite{shuman13-mag}. 

Consider the signal $\bbx=[\bbx^1;\ldots;\bbx^F]$ formed by $F$ feature vectors $\bbx^f$ with $N$ components each. The feature vector $\bbx^f$ is said to be a graph signal when each of its $N$ components is assigned to a different vertex of the graph. The graph describes the underlying support of the data $\bbx^f$ (hence, of $\bbx$) by using the weights $\ccalW$ to encode arbitrary pairwise relationships between data elements. The graph shift enables processing of the graph signal $\bbx^f$ because it defines a local linear operation that can be applied to graph signals. Indeed, if we consider the signal $\bby^f:=\bbS\bbx^f$ it follows from the sparsity of $\bbS$ that the $n$th element of $\bby^f$ depends on the elements of $\bbx^f$ associated with neighbors of the node $n$,
\begin{equation}\label{eqn_gso_is_a_shift}
   [\bby^f]_n =  \sum_{m:(m,n)\in\ccalE} [\bbS]_{nm} [\bbx^f]_m.
\end{equation}
It is instructive to consider the cyclic graph adjacency matrix $\bbA_{\dc}$, with nonzero elements $[\bbA_{\dc}]_{1 + n \mod N, n}=1$. Since the cyclic graph describes the structure of discrete (periodic) time, we can say that a discrete time signal $\bbx$ is a graph signal defined on the cyclic graph. When particularized to $\bbS=\bbA_{\dc}$, \eqref{eqn_gso_is_a_shift} yields ${y^f_{1+n\mod N}}=x^f_n$ implying that $\bby^f$ is a circularly time shifted copy of $\bbx^f$. This motivates interpretation of $\bbS$ as the generalization of time shifts to signals supported in the corresponding graph \cite{sandryhaila13-dspg}.

Enabling CNNs to process data modeled as graph signals entails extending the operations of convolution and pooling to handle the irregular nature of the underlying support. Convolution [cf.~\eqref{eqn:conv_time}] can be readily replaced by the use of linear, shift invariant graph filters [cf.~\eqref{eqn:conv_graph}]. The summarizing function [cf.~\eqref{eqn_group_nonlinearity}] can also be readily extended by using the notion of neighborhood defined by the underlying graph support. The pointwise nonlinearity can be kept unmodified [cf.~\eqref{eqn_downsampling}], but there are two general downsampling strategies for graph signals: selection sampling \cite{chen15-selection} and aggregation sampling \cite{marques16-aggregation}. Inspired by these, we propose two architectures: selection GNNs (Section \ref{sec:selection}) and aggregation (Section \ref{sec:aggregation}) GNNs.

%
\begin{remark} \normalfont 
Although our current theoretical understanding of CNNs is limited, empirical evidence suggests that convolution and pooling work in tandem to act as feature extractors at different levels of resolution. At each layer, the convolution operation linearly relates up to $K_{\ell}$ nearby values of each input feature. Since the same filter taps are used to process the whole signal, the convolution finds patterns that, albeit local, are irrespective of the specific location of the pattern in the signal. The use of several features allows collection of different patterns through learning of different filters thus yielding a more expressive operation. The pooling stage summarizes information into a feature of lower dimensionality. It follows that subsequent convolutions operate on summaries of different regions. As we move into deeper layers we pool summaries of summaries that are progressively growing the region of the signal that affects a certain feature. The conjectured value of composing local convolutions with pooling summaries is adopted prima facie as we seek graph neural architectures that exploit the locality of the shift operator to generalize convolution and pooling operations.
\end{remark}


\section{Selection Graph Neural Networks} \label{sec:selection}



Generalizing the first layer of a CNN to signals supported on graphs is straightforward as it follows directly from the definition of a linear shift invariant filter \cite{segarra17-linear}. Going back to the definition of convolutional features in \eqref{eqn:conv_time} we reinterpret the filters $\bbh_{1}^{fg}$ as graph filters that process the features $\bbx_0^g$ through a graph convolution. This results in intermediate features $\bbu_{1}^{fg}$ having components 
\begin{equation} \label{eqn:conv_graph}
   \Big[\bbu_{1}^{fg}\Big]_n
      := \Big[\bbh_{1}^{fg} \ast_\bbS \bbx_0^g \Big]_{n} 
      := \sum_{k=0}^{K_1-1} 
             \Big[ \bbh_{1}^{fg} \Big]_{k} 
             \Big[ \bbS^{k} \bbx_{0}^{f}\Big]_{n},
\end{equation}
where we have used $\ast_\bbS$ to denote the graph convolution operation on $\bbS$. The summations in equations \eqref{eqn:conv_time} and \eqref{eqn:conv_graph} are analogous except for the different interpretations of what it means to shift the input signal $\bbx_{0}^{f}$. In \eqref{eqn:conv_time}, a $k$-unit shift at index $n$ means considering $[\bbx_{0}^{f}]_{n-k}$, the value of the signal $\bbx_{0}^{f}$ at time $n-k$. In \eqref{eqn:conv_graph}, graph shifting at node $n$ entails the operation $[ \bbS^{k} \bbx_{0}^{f}]_{n}$ which composes a multiplication by $\bbS^k$ with the selection of the resulting value at $n$. In fact, particularizing \eqref{eqn:conv_graph} to the cyclic graph by making $\bbS=\bbA_{\dc}$ renders \eqref{eqn:conv_time} and \eqref{eqn:conv_graph} equivalent. From the perspective of utilizing \eqref{eqn:conv_graph} as an extractor of local (graph) convolutional features it is important to note that graph convolutions aggregate information through successive local operations [cf. \eqref{eqn_gso_is_a_shift}]. A filter with $K_1$ taps incorporates information at node $n$ that comes from nodes in its $(K_1-1)$-hop neighborhood.

%
\begin{figure*}[t]
\begin{center}
\begin{tabular}{lll}
\includegraphics[width=0.3\textwidth]{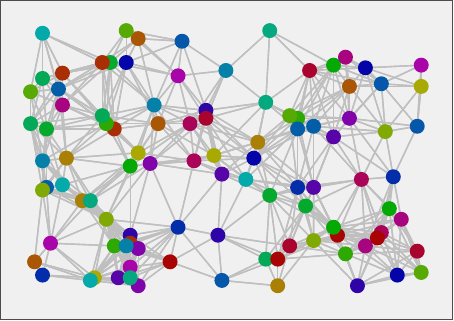} & 
\includegraphics[width=0.3\textwidth]{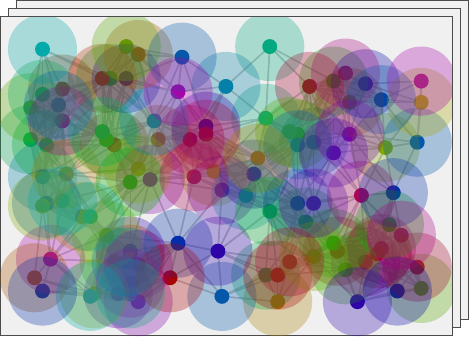}  &
\includegraphics[width=0.3\textwidth]{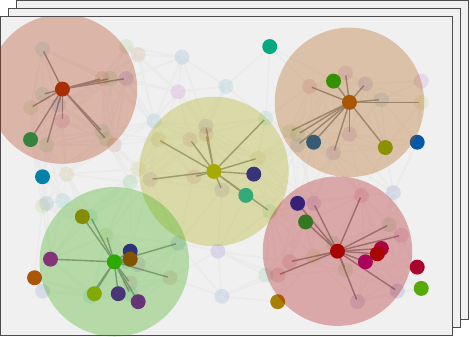}  \\ \\ 
\includegraphics[width=0.3\textwidth]{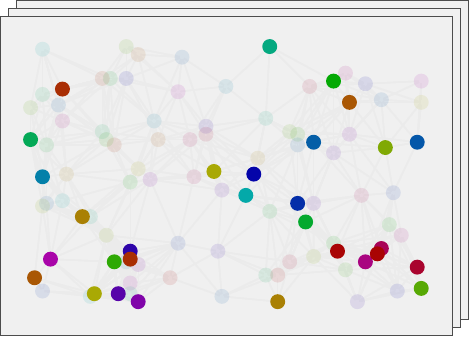} &
\includegraphics[width=0.3\textwidth]{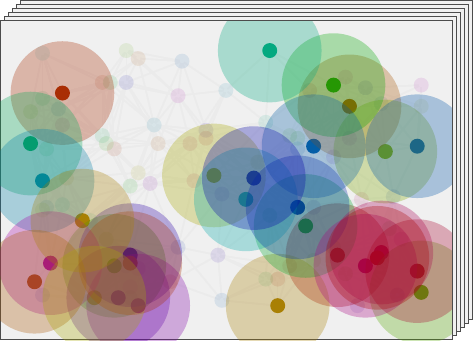}  & 
\includegraphics[width=0.3\textwidth]{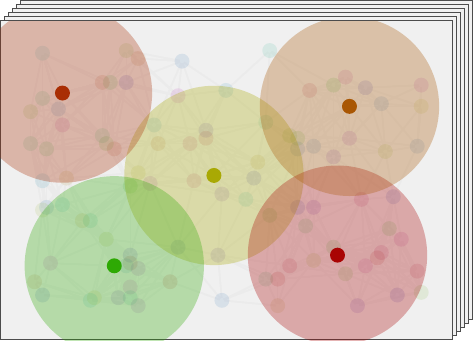}  \\ \\ 
\includegraphics[width=0.3\textwidth]{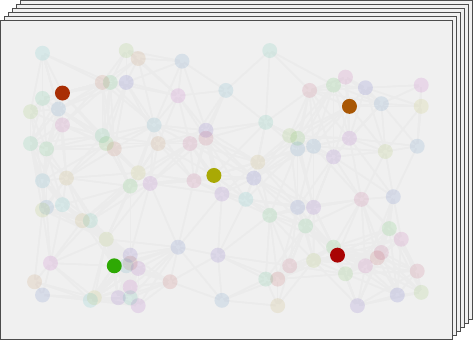} & 
\includegraphics[width=0.3\textwidth]{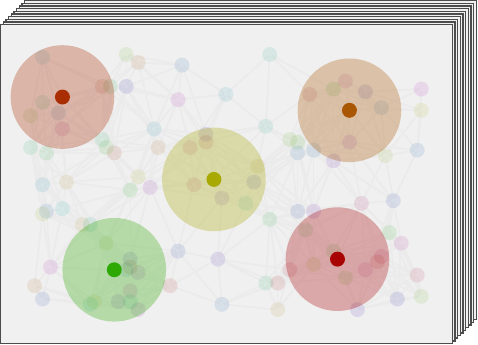}  & 
\includegraphics[width=0.3\textwidth]{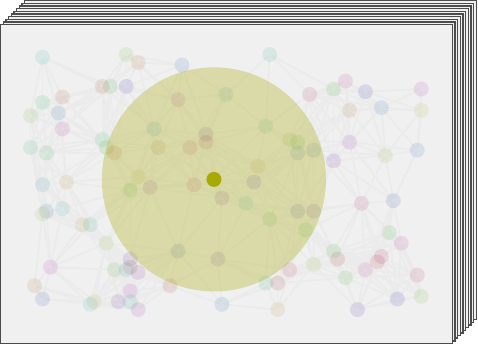}
\end{tabular}
\end{center}  	
\caption{Selection Graph Neural Networks. Consider the input to be a signal supported by a known $N$-node graph. First, convolutional features are obtained by means of graph filtering in the original graph [cf.~\eqref{eqn:graph_filter}]. The color disks in the second column illustrate the convolution operation on each node. Then, a subset of $N_{1}$ nodes is selected, and summarizing function $\rho_{1}$ and pointwise nonlinearity $\sigma_{1}$ are applied to the neighborhood $\bbn_{1}$ for each of these nodes, obtaining the output $\bbx_{1}^{f}$ for the first layer. The color disks in the third column show the reach of the pooling operation, the size of the neighborhood being pooled (in the first row, the disks include only the one-hop neighborhood; also, only a few disks are shown so as not to clutter the illustration). In order to obtain convolutional features for following layers, we zero pad the signal to \emph{fit} the original graph [cf.~\eqref{eqn_zero_padding}] so as to apply a graph filter and then resample the output at the same set of nodes [cf.~\eqref{eqn_conv_features_unpadded_preliminary}-\eqref{eqn_conv_features_unpadded}]. Then, a new smaller subset of nodes is selected, and the summarizing function and pointwise nonlinearity are applied to a neighborhood of these nodes [cf. \eqref{eqn_graph_neigborhood}]. This process is repeated while selecting fewer and fewer nodes.}
\label{fig:selection_cnn}
\end{figure*}

Although we wrote \eqref{eqn:conv_graph} componentwise to emphasize its similarity with \eqref{eqn:conv_time} we can drop the $n$ subindices to write a vector relationship. For future reference we further define the linear shift invariant filter $\bbH_{1}^{fg} := \sum_{k=0}^{K_1-1} [\bbh_{1}^{fg}]_{k} \bbS^{k}$ to write
\begin{equation} \label{eqn:graph_filter}
   \bbu_{1}^{fg} 
	\ =  \ \sum_{k=0}^{K_1-1} 
	             \Big[ \bbh_{1}^{fg} \Big]_{k} \,
	             \bbS^{k} 
	             \bbx_{0}^{f}
    \ := \ \bbH_{1}^{fg}\bbx_{0}^{f}.
\end{equation}
The graph filter \eqref{eqn:graph_filter} is a generalization of the Chebyshev filter in \cite{defferrard17-cnngraphs}. More precisely, if $\ccalG$ is an undirected graph, and we adopt the normalized Laplacian as the graph shift operator $\bbS$, then \eqref{eqn:graph_filter} boils down to a Chebyshev filter. The convolutional stage in \cite{kipf17-classifgcnn} is a Chebyshev filter of $K=2$, and thus is also a special case of \eqref{eqn:graph_filter}. We also note that the use of polynomials on arbitrary graph shift operators for the convolutional stage has been also proposed in \cite{du17-topoadapt, gama18-nvgf}.
Asides from replacing the linear time invariant filter in \eqref{eqn:conv_time} with the graph shift invariant filter in \eqref{eqn:graph_filter}, the remaining components of the conventional CNN architecture can remain more or less unchanged. The feature aggregation in \eqref{eqn:agg_features} to obtain $\bbu_{1}^{f}$ needs no modification as it is a simple summation independent of the graph structure. The summarization operator in \eqref{eqn_group_nonlinearity} requires a redefinition of locality. This is not difficult because it follows from \eqref{eqn:graph_filter} that $\bbu_{1}^{f}$ is another $N$-node graph signal that is defined over the same graph $\bbS$. We can then use $\bbn_1$ to represent a graph neighborhood of node $n$ and apply the same summary operator. We point out that $\bbn_1$ need not be the 1-hop neighborhood of $n$. The sampling and activation operation in \eqref{eqn_downsampling} requires a matrix $\bbC_1$ to sample over the irregular graph domain. Apart from the challenge of selecting sampling matrices for graphs -- see \eqref{eqn:selection_set} and \cite{chen15-selection, marques16-aggregation, anis16-spectralproxies, tsitsvero16-uncertainty} --, this does not require any further modification to \eqref{eqn_downsampling}. The first row of Fig.~\ref{fig:selection_cnn} shows the operations carried out in this first layer.

The challenge in generalizing CNNs to GNNs arises beyond the first layer. After implementing the sampling operation in \eqref{eqn_downsampling} the signal $\bbx^{f}_{1}$ is of lower dimensionality than $\bbu_{1}^{f}$ and can no longer be interpreted as a signal supported on $\bbS$. In regular domains this is not a problem because we use the extraneous geometrical information of the underlying domain to define convolutions in the space of lower dimensionality. To see this in terms of graph signals, let us interpret the signal $\bbx_{0}^{g}$ defined on a regular domain as one defined on a cyclic graph with $N_0=N$ nodes, which is also the same graph that describes  $\bbu_{1}^{f}$. Then, if we consider a downsampling factor of $(N_{1}/N_{0})$,  another \textit{cyclic} graph with $N_1$ nodes describes the signal $\bbx_{1}^{f}$. However, when graph signals are defined in a generic irregular domain, there is no extraneous information to elucidate the form of the graph that describes signals beyond the first layer. Resolving mismatched supports is a well-known problem in signal processing whose simplest and most widely-used solution is zero padding. The following sections illustrate how zero padding can be leveraged to resolve one of the critical challenges in the implementation of GNNs.

%
\subsection{Selection Sampling on Graph Convolutional Features}\label{sec_selection_convolution}

Sampling is an operation that selects components of a signal. To explain the construction of convolutional features on graphs, it is more convenient to think of sampling as the \textit{selection of nodes} of a graph which we call active nodes. This implies that at each layer $\ell$ we place the input features $\bbx^f_{\ell-1}$ of dimension $N_{\ell-1}$ on top of the active nodes of the graph $\bbS$. Selection schemes are further discussed in Sec.~\ref{subsec_selection_practical}. Doing so requires that we keep track of the location of the samples. Thus, at each layer $\ell$ we consider input features $\bbx_{\ell-1}^{g}$ each with $N_{\ell-1}$ components, and zero padded features $\tbx_{\ell-1}^{g}$ each with size $N$ but only $N_{\ell-1}$ nonzero components which replicate the values of $\bbx_{\ell-1}^{g}$. The indexes of the nonzero components of $\tbx_{\ell-1}^{g}$ correspond to the location of the elements of $\bbx_{\ell-1}^{g}$ in the original graph. It is clear that we can move from the unpadded to the padded representation by multiplying with an $N \times N_{\ell-1}$ tall binary sampling matrix $\bbD_{\ell-1}^{\Tr}$. Indeed, if we let $[\bbD_{\ell-1}]_{mn}=1$ represent the $m$th component of the unpadded feature, $[\bbx_{\ell-1}^{g}]_m$, is located in the $n$th node of the graph, we can write the padded feature as
\begin{equation} \label{eqn_zero_padding}
   \tbx_{\ell-1}^{g} \ =\ \bbD_{\ell-1}^{\Tr} \bbx_{\ell-1}^{g} .
\end{equation}
The advantage of keeping track of the padded signal is that convolutional features can be readily obtained by operating in the original graph. Given the notion of graph convolution in \eqref{eqn:graph_filter} and (re-)defining $\bbh_{\ell}^{fg}$ to be the graph filter coefficients at layer $\ell$ we can define intermediate features as [cf. \eqref{eqn:conv_time}]
\begin{equation} \label{eqn_conv_features_padded}
   \tbu_{\ell}^{fg} 
	\ :=  \ \sum_{k=0}^{K_{\ell}-1} 
	             \Big[ \bbh_{\ell}^{fg} \Big]_{k} \,
	             \bbS^{k} \,
	             \tbx_{\ell-1}^{g} .
\end{equation}
Although a technical solution to the construction of convolutional features, \eqref{eqn_conv_features_padded} does not exploit the computational advantages of sampling. These can be recovered by selecting components of $\tbu_{\ell}^{fg}$ at the same set of nodes that support $\bbx_{\ell-1}^{g}$. We then define $\bbu_{\ell}^{fg} := \bbD_{\ell-1} \tbu_{\ell}^{fg}$. If we further use \eqref{eqn_zero_padding} to substitute $\tbx_{\ell-1}^{g}$ into the definition of the convolutional features in \eqref{eqn_conv_features_padded}, we can write
\begin{equation} \label{eqn_conv_features_unpadded_preliminary}
   \bbu_{\ell}^{fg}
     := \, \bbD_{\ell-1} \tbu_{\ell}^{fg} 
	  = \, \bbD_{\ell-1}
	       \sum_{k=0}^{K_{\ell}-1} 
	               \Big[ \bbh_{\ell}^{fg} \Big]_{k} \,
	               \bbS^{k} \,
	               \bbD_{\ell-1}^{\Tr} \,
	               \bbx_{\ell-1}^{g} .
\end{equation}
If we further define reduced dimensionality $k$-shift matrices 
\begin{equation} \label{eqn_gsp_unpadded_powers}
   \bbS_{\ell}^{(k)} :=  \bbD_{\ell-1}\, \bbS^{k}\,\bbD_{\ell-1}^{\Tr},
\end{equation}
and reorder and regroup terms in \eqref{eqn_conv_features_unpadded_preliminary} we can reduce the definition of convolutional features to 
\begin{equation} \label{eqn_conv_features_unpadded}
   \bbu_{\ell}^{fg}
	\ =  \ \sum_{k=0}^{K_{\ell}-1} 
	             \Big[ \bbh_{\ell}^{fg} \Big]_{k} \,
	             \bbS_{\ell}^{(k)}\,
	             \bbx_{\ell-1}^{g} 
	\:=  \ \bbH_{\ell}^{fg} \bbx_{\ell-1}^{g} ,
\end{equation}
where we have also defined the \textit{subsampled} linear shift invariant filter $\bbH_{\ell}^{fg} :=  \sum_{k=0}^{K_{\ell}-1} [\bbh_{\ell}^{fg}]_{k}\bbS_{\ell}^{(k)}$. Implementing \eqref{eqn_conv_features_unpadded_preliminary} entails repeated application of the shift operator to the padded signal, which can be carried out with low cost if the original input graph is sparse. In \eqref{eqn_conv_features_unpadded}, the filter $\bbH_{\ell}^{fg}$ takes advantage of sampling to operate directly on a space of lower dimension $N_{\ell-1}$. The matrices $\bbS_{\ell}^{(k)}$ can be computed beforehand because they depend on the graph shift operator and the sampling matrices only. We emphasize that, save for subsampling, \eqref{eqn_conv_features_unpadded} and \eqref{eqn_conv_features_unpadded_preliminary} are equivalent and that, therefore, the features $\bbu_{\ell}^{fg}$ generated by the subsampled filter $\bbH_{\ell}^{fg}$ are convolutional relative to the original graph (shift) $\bbS$. The middle image in Fig.~\ref{fig:selection_cnn} shows zero pad of input signal, convolution in the original graph, and resampling of the filter output.

Features $\bbu_{\ell}^{f}$ can be obtained from features $\bbu_{\ell}^{fg}$ using the same linear aggregation operation in \eqref{eqn:agg_features} which does not require adaptation to the structure of the graph,
\begin{equation} \label{eqn_agggregated_features_graph}
   \bbu_{\ell}^{f}
	   \  =\ \sum_{g=1}^{F_{\ell-1}} \bbH_{\ell}^{fg} \bbx_{\ell-1}^{g}.
\end{equation}
This completes the construction of convolutional features and leads to the pooling stage we describe next.

%
\subsection{Selection Sampling and Pooling}\label{sec_selection_pooling}

The pooling stage requires that we redefine the summary and sampling operations in \eqref{eqn_group_nonlinearity} and \eqref{eqn_downsampling}. Generalizing the summary operation requires redefining the aggregation neighborhood. In the first layer, this can be readily accomplished by selecting the $\alpha_{1}$-hop neighborhood of each node for some given $\alpha_{1}$ that defines the reach of the summary operation. This information is actually contained in the powers of the shift operator. The 1-hop neighborhood of $n$ is the set of nodes $m$ such that $[\bbS]_{nm} \neq 0$, the 2-hop neighborhood is the union of this set with those nodes $m$ with $[\bbS^2]_{nm} \neq 0$ and so on. In the case of the sampled features the graph neighborhoods need to be intersected with the set of active nodes. This intersection is already captured by the $k$-shift matrices $\bbS_{\ell}^{(k)}$ [cf. \eqref{eqn_gsp_unpadded_powers}]. Thus, at layer $\ell$ we introduce an integer $\alpha_{\ell}$ to specify the reach of the summary operator and define the $\alpha_{\ell}$-hop neighborhood of $n$ as
\begin{equation} \label{eqn_graph_neigborhood}
   \bbn_\ell = \left[ m: \big[\bbS_{\ell}^{(k)}\big]_{nm} \neq 0,\ 
                       \text{for some}\ k\leq \alpha_{\ell} \right].
\end{equation}
Summary features $[\bbv^{f}_{\ell}]_n$ at node $n$ are computed from \eqref{eqn_group_nonlinearity} using the graph neighborhoods in \eqref{eqn_graph_neigborhood}. These neighborhoods follow the node proximity encoded by $\bbS$, see third column of Fig.~\ref{fig:selection_cnn}.

To formally explain the downsampling operation in \eqref{eqn_downsampling} in the context of graph signals, begin by defining sampling matrices adapted to irregular domains. This can be easily defined at the $\ell$th  layer if we let the sampling matrix $\bbC_{\ell}$ be a fat matrix with $N_{\ell}$ rows and $N_{\ell-1}$ columns with the properties
\begin{equation} \label{eqn:selection_set}
   [\bbC_{\ell}]_{mn} \in \{0,1\}       , \quad
   \bbC_{\ell} \bbone = \bbone          , \quad
   \bbC_{\ell}^{\Tr} \bbone \leq \bbone .
\end{equation}
When $[\bbC_{\ell}]_{mn}=1$ it means that the $n$th component of $\bbv^{f}_{\ell}$ is selected in the product $\bbC_{\ell}\bbv^{f}_{\ell}$ and stored as the $m$th component of the output. The properties in \eqref{eqn:selection_set} ensure that exactly $N_{\ell}$ components of $\bbv^{f}_{\ell}$ are selected and that no component is selected more than once. They do not, however, enforce a regular sampling pattern. We further define the nested sampling matrix $\bbD_{\ell}$ as the product of all sampling matrices applied up until layer $\ell$
\begin{equation}\label{eqn:nested_sampling_matrices}
   \bbD_{\ell}\ =\ \bbC_{\ell} \bbC_{\ell-1}  \ldots  \bbC_{1}
              \ =\  \prod_{\ell'=1}^\ell \bbC_{\ell'} .
\end{equation}
Matrix $\bbD_{\ell}$ keeps track of the location of the selected nodes in the original graph, for each layer $\ell$, and is thus used for the zero padding operation in \eqref{eqn_conv_features_unpadded_preliminary}.

Each layer of the selection GNN architecture is determined by \eqref{eqn_conv_features_unpadded}-\eqref{eqn_agggregated_features_graph} for the convolution operation and \eqref{eqn_group_nonlinearity}-\eqref{eqn_downsampling} for pooling and nonlinearity. To summarize, the input to layer $\ell$ is $\bbx_{\ell-1}$ comprised of $F_{\ell-1}$ features $\bbx_{\ell-1}^{f}$ located at a subset of nodes given by $\bbD_{\ell-1}$. Then, we use the reduced dimensionality $k$-shift matrices \eqref{eqn_gsp_unpadded_powers} to process $\bbx_{\ell-1}^{f}$ using a graph filter as in \eqref{eqn_conv_features_unpadded}, and obtain aggregated features $\bbu_{\ell}^{f}$ \eqref{eqn_agggregated_features_graph}. A neighborhood $\bbn_{\ell}$ for each element of $\bbu_{f}$ is determined by \eqref{eqn_graph_neigborhood} for some $\alpha_{\ell}$ and the output $\bbv_{\ell}^{f}$ of the summarizing function $\rho_{\ell}$ is computed as in \eqref{eqn_group_nonlinearity}. Finally, following \eqref{eqn_downsampling}, a smaller subset of nodes is selected by means of $\bbC_{\ell}$ and the pointwise nonlinearity $\sigma_{\ell}$ is applied to obtain the $\ell$th output features $\bbx_{\ell}^{f}$, for $f=1,\ldots,F_{\ell}$. See Algorithm~\ref{algm_selection_gnn} for details.

\begin{algorithm}[t]
 	\caption{Selection Graph Neural Network.}
	\label{algm_selection_gnn}

	\begin{algorithmic}[1]
 \Statex \textbf{Input:} $\{\hbx\}$: testing dataset, $\ccalT$: training dataset
 \Statex $\quad \bbS$: graph shift operator, $L$: Number of layers, 
 \Statex $\quad \{F_{\ell}\}$: number of features, $\{K_{\ell}\}$: degree of filters
 \Statex $\quad \{\rho_{\ell}\}$: neighborhood summarizing function
 \Statex $\quad \texttt{selection}$: selection sampling method
 \Statex $\quad \{N_{\ell}\}$: number of nodes on each layer
 \Statex $\quad \{\sigma_{\ell}\}$: pointwise nonlinearity
 \Statex \textbf{Output:} $\{\hby\}$: estimates of $\{\hbx\}$
 \Statex
 \Procedure{selection\_gnn}{$\{\hbx\}$, $\ccalT$, $\bbS$, $L$, $\{F_{\ell}\}$,$\{K_{\ell}\}$, $\{\rho_{\ell}\}$, $\texttt{selection}$, $\{N_{\ell}\}$, $\{\sigma_{\ell}\}$}
 {\Statex{$\rhd$ \emph{Create architecture:}}}
   \For{$\ell = 1:L-1$}
       \State Compute $\bbD_{\ell-1} = \bbC_{\ell-1} \bbD_{\ell-2}$ \Comment{\emph{See \eqref{eqn:nested_sampling_matrices}}}
       \State Compute $\bbS_{\ell}^{(k)}$ for $k=0,\ldots,K_{\ell}-1$ \Comment{\emph{See \eqref{eqn_gsp_unpadded_powers}}}
       \State Create $[\bbh_{\ell}^{fg}]_{k}$, $f = 1,\ldots,F_{\ell}$, $g=1,\ldots,F_{\ell-1}$
       \State Compute filters $\bbH_{\ell}^{fg} = \sum_{k=0}^{K_{\ell-1}} [\bbh_{\ell}^{fg}]_{k} \bbS_{\ell}^{(k)}$
       \State Aggregate filtered features $\sum_{g=1}^{F_{\ell-1}} (\bbH_{\ell}^{fg} \cdot)$
       \State Apply summarizing function $\rho_{\ell}(\cdot)$
       \State Select $N_{\ell}$ nodes following method $\texttt{selection}$
       $$ \bbC_{\ell} = \texttt{selection}(N_{\ell},\bbC_{\ell-1}) $$
       \State Downsample output of summarizing function $\bbC_{\ell} \rho_{\ell}$
       \State Apply pointwise nonlinearity $\sigma_{\ell}(\cdot)$
   \EndFor
   \State Create fully connected layer $\bbA_{L} \cdot$
 \Statex{$\rhd$ \emph{Train:}}
   \State Learn $\{[\bbh_{\ell}^{fg}]_{k}\}$ and $\bbA_{L}$ from $\ccalT$
 \Statex{$\rhd$ \emph{Evaluate:}}
   \State Obtain $\hby$ applying GNN on $\hbx$ with learned coefficients
 \EndProcedure
 	\end{algorithmic}

\end{algorithm}

\begin{remark}\normalfont
The selection GNN architecture recovers a conventional CNN when particularized to graph signals described by a cyclic graph (conventional discrete time signals). To see this, let $\bbS=\bbA_{\dc}$ for a graph of size $N$, and let $\bbC_{\ell-1}$ be the sampling matrix that takes $N_{\ell-1}$ equally spaced samples out of the previous $N_{\ell-2}$ samples, for every $\ell$. Then, the nested sampling matrix $\bbD_{\ell-1}$ becomes a sampling matrix that takes $N_{\ell-1}$ equally spaced samples out of the $N$ original ones. This implies that $\bbS_{\ell}^{(k)} = \bbD_{\ell-1} \bbA_{\dc}^{k} \bbD_{\ell-1}^{\Tr}$ becomes either the $k$th power of the adjacency matrix of a cyclic graph with $N_{\ell-1}$ nodes for $k$ a multiple of $N/N_{\ell-1}$, or the all-zero matrix otherwise. This results in convolutional features obtained by \eqref{eqn_conv_features_unpadded} being equivalent to those obtained by \eqref{eqn:conv_time}. Likewise, making $\alpha_{\ell}=N_{\ell-1}/N_{\ell}$ for all $\ell$ leads to regular pooling and downsampling. This shows that the selection GNN does indeed boil down to the conventional CNN for discrete time signals.
\end{remark}

\begin{remark}\normalfont
The dimension $N_{\ell}$ is being effectively reduced without the need to use a complex multiscale hierarchical clustering algorithm. More specifically, in each layer, only a new set of nodes is used, but there is no need to recompute edges between these nodes or new weight functions, since the underlying graph on which the operations are actually carried out is the same graph support as the initial input data $\bbx$. This, not only avoids the computational cost of obtaining multiscale hierarchical clusters, but also avoids the need to assess when such clustering scheme is adequate.
\end{remark}

%
\subsection{Practical Considerations} \label{subsec_selection_practical}

\myparagraph{Selection of nodes.} There is a vast GSP literature on sampling by selecting nodes, see, e.g., \cite{chen15-selection, anis16-spectralproxies, tsitsvero16-uncertainty, puy15-scores, varma15-scores}. In this paper, we consider that any one of these methods is adopted throughout the Selection GNN, and at each layer $\ell$ matrix $\bbC_{\ell}$ is determined by following the chosen method. On each layer $\ell$ the subset of nodes selected by $\bbC_{\ell}$ is always a subset of the nodes chosen in the previous layer. This implies that $N_{\ell} \leq N_{\ell-1}$ and that $\bbC_{\ell} \bbC_{\ell-1}$ never yields the zero matrix. In particular, in Sec.~\ref{sec:sims}, we adopt the methods proposed in \cite{anis16-spectralproxies} and \cite{varma15-scores} to study their impact on the overall performance of the Selection GNN.

\myparagraph{Locality of filtering.} The graph convolution remains a local operation with respect to the original input graph. Since each convolutional feature is zero padded to fit the graph, the implementation of the graph filter at each layer can be carried out by means of local exchanges in the original support. This can be a good computational option if the original input graph is sparse, and therefore repeatedly applying the graph shift operator exploits this sparsity. This turns out to be particularly useful when such a support represents a physical network with physical connections.

\myparagraph{Centralized computing.} When regarding the selection pooling architecture as a whole, being executed from a single centralized unit (i.e. when local connectivity is not important for computation purposes, for example, in the training phase), it is observed that the computational cost of carrying out convolutions \eqref{eqn_conv_features_unpadded} is reduced to matrix multiplication in the smaller $N_{\ell}$-dimensional space. It is noted that the reduced dimensionality $k$-shift matrices \eqref{eqn_gsp_unpadded_powers} can be obtained before the training phase, and also, that the statistical properties of learning the filter taps are not affected by it. This observation, coupled with the previous one, shows that the selection pooling architecture adequately addresses the global vs. local duality by efficiently computing convolutions in both settings.

\myparagraph{Computation of nonlinearities.} From an implementation perspective, it is observed that, while the local summarizing function $\rho_{\ell}$ involves the neighborhood of the $N_{\ell-1}$ nodes (which are more than the $N_{\ell}$ nodes that are kept in layer $\ell$), this function only has to be computed for those $N_{\ell}$ nodes that are left after downsampling. That is, it is not needed to compute $\rho_{\ell}$ at each one of the $N_{\ell-1}$ nodes, but only at the $N_{\ell}$ nodes that are actually kept after downsampling. In this sense, this nonlinear operation can be subsumed with the pointwise nonlinearity $\sigma_{\ell}$ that is applied to the $N_{\ell}$ nodes. To further illustrate this point, suppose that max-pooling is used and that the corresponding pointwise nonlinearity is a ReLU, $\sigma_{\ell}(x) = \max\{0,x\}$. Then, both operations can be performed simultaneously at node $n$ by doing $\max\{0,\{x_{m} : (m,n) \in \bbn_{\ell}\}\}$, where $\bbn_{\ell}$ denotes the paths in the neighborhood, and where this operation is computed only for nodes $n$ that are part of the $N_{\ell}\leq N_{\ell-1}$ selected nodes.

\myparagraph{Regularization of filter taps.} As the Selection GNN grows in depth (more layers), the number of filter taps in the convolution stage might increase, in order to access information located at further away neighbors (this happens if the few selected nodes at some deeper layer are far away from each other, as measured by the number of neighborhood exchanges). It is a good idea, then, to structure the filter coefficients $\bbh_{\ell}^{fg}$ in these deeper layers. More specifically, filtering with $N$ taps might be necessary, so it makes sense to choose $[\bbh_{\ell}^{fg}]_k$ constant for a range of $k$, since no new substantial information is going to be included for a wide range of those $k$. This reduces the number of trainable parameters and consequently overfitting.

\myparagraph{Definition of neighborhoods.} Information from the weight function $\ccalW$ of the graph can be included in the pooling stage \eqref{eqn_graph_neigborhood}. More precisely, instead of defining the neighborhood only looking at the edge set $\ccalE$, that is $[\bbS_{\ell}^{(k)}]_{nm} \neq 0$, we can make $[\bbS_{\ell}^{(k)}]_{nm} \geq \delta$ so that we summarize only across edges stronger than $\delta$.

\myparagraph{Frequency interpretation of convolutional features.} One advantage of having convolutional features defined always on the same graph $\ccalG$ at every layer $\ell$ is that these can be easily analyzed from a frequency perspective. Since the graph Fourier transform of a signal depends on the eigenvectors $\bbV$ of the graph shift operator \cite{sandryhaila14-freq}, and since the same $\bbS = \bbV \bbLambda \bbV^{-1}$ is used to define all convolutional features [cf. \eqref {eqn_conv_features_unpadded_preliminary}], then they all share the same frequency basis, allowing for a comprehensive frequency analysis at all layers. In particular, if we focus on normal matrix GSOs, i.e. $\bbV^{-1} = \bbV^{\Hr}$, the zero-padding aliasing effect is evidenced in the fact that $\bbV^{\Hr} \bbD^{\Tr} \bbD \bbV$ need not be the identity matrix for arbitrary eigenvectors $\bbV$ and downsampling matrices $\bbD$, altering the frequency content of the input signal to a filter. However, the filter taps are learned from the training set, taking into account this aliasing effect, and therefore are able to cope with it, extracting useful features.

\myparagraph{Computational cost.} The number of computations at each layer is given by the cost of the convolution operation, which is $O(|\ccalE|K_{\ell} F_{\ell} F_{\ell-1})$ if \eqref{eqn_conv_features_unpadded_preliminary} is used, or $O(N_{\ell-1}^{2} K_{\ell} F_{\ell} F_{\ell-1})$ if \eqref{eqn_conv_features_unpadded} is used, since pooling and downsampling incur in negligible cost. We observe that in \eqref{eqn_conv_features_unpadded} the cost tends to be dominated by $N_{\ell-1}^{2}$ making dimensionality reduction (i.e. pooling) a critical step for scalability.

%
\begin{figure*}[!t]
\centering
\includegraphics[width=0.2300\textwidth]{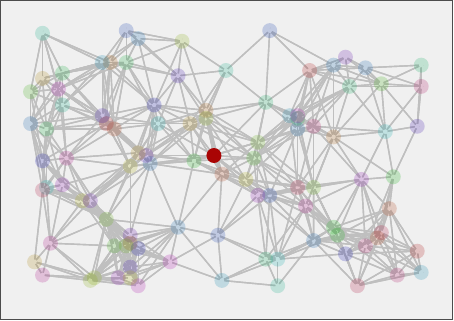} \quad 
\includegraphics[width=0.2300\textwidth]{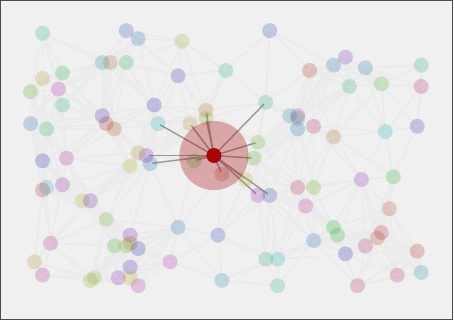} \quad
\includegraphics[width=0.2300\textwidth]{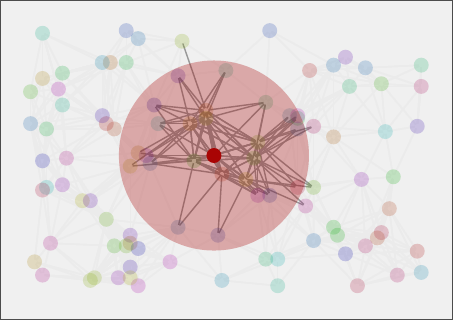} \quad
\includegraphics[width=0.2300\textwidth]{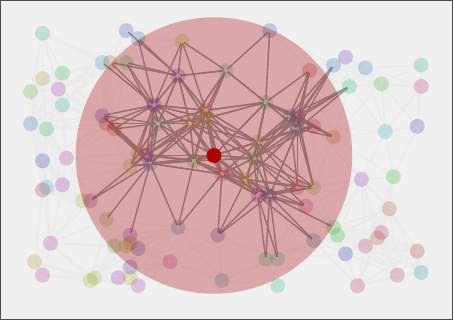} \\ \vspace{2mm}
\includegraphics[width=1.0000\textwidth]{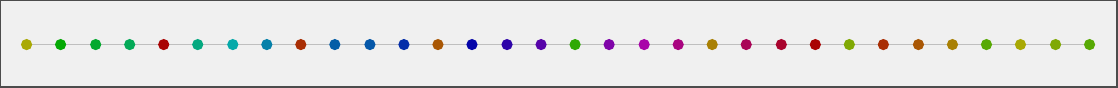}  \\ \vspace{2mm}
\includegraphics[width=1.0000\textwidth]{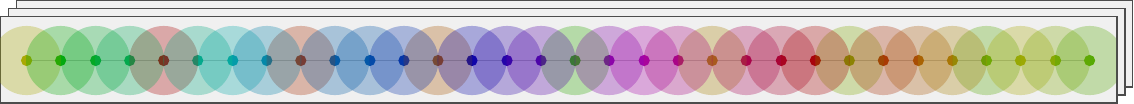}   \\ \vspace{2mm}
\includegraphics[width=1.0000\textwidth]{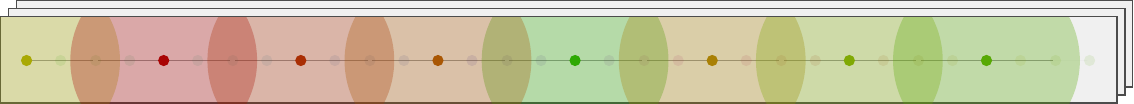}   \\ \vspace{2mm}
\includegraphics[width=1.0000\textwidth]{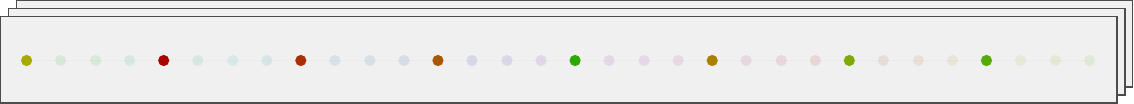} 
\caption{Aggregation Graph Neural Networks. Select a node $p \in \ccalV$ and perform successive local exchanges with its neighbors. For each $k$-hop neighborhood (illustrated by the increasing disks in the first row), record $\bbS^{k}\bbx^{g}$ at node $p$ and build signal $\bbz_{p}^{g}$ which exhibits a regular structure [cf.~\eqref{eqn:agg_representation}]. Once a regular time-structure signal is obtained, we proceed to apply regular convolution and pooling to process the data [cf. \eqref{eqn:conv_time}-\eqref{eqn_downsampling}].}
	\label{fig_aggregation}
\end{figure*}

\myparagraph{Number of parameters.} The number of parameters to be learned at each layer are determined by the length of the filters, and the number of input and output features and is given by $O(K_{\ell} F_{\ell} F_{\ell-1})$ independent of $N_{\ell-1}$.


\section{Aggregation Graph Neural Networks} \label{sec:aggregation}


%
The {\it selection} GNNs of Section \ref{sec:selection} create convolutional features adapted to the structure of the graph with linear shift invariant graph filters. The {\it aggregation} GNNs that we describe here apply the conventional CNN architecture of Section \ref{sec:regular} to a signal with temporal (regular) structure that is generated to incorporate the topology of the graph. To create such a temporal arrangement we consider successive applications of the graph shift operator $\bbS$ to the input graph signal $\bbx^g$ (see first row of Fig.~\ref{fig_aggregation}). This creates a sequence of $N$ graph shifted signals $\bby^g_0, \ldots, \bby^g_{N-1}$. The first signal of the sequence is $\bby^g_0 = \bbx^g$, the second signal is $\bby^g_1 = \bbS\bbx^g$, and subsequent members of the sequence are recursively obtained as $\bby^g_k = \bbS\bby^g_{k-1} = \bbS^{k}\bbx^g$. We observe that each vector $\bby_{k}^{g}$ incorporates the underlying support by means of multiplication by the graph shift operator $\bbS$. We arrange the sequence of signals $\bby^g_k$ into the matrix representation of the graph signal $\bbx^g$ that we define as
\begin{equation} \label{eqn:aggregation_matrix}
   \,\, \bbX^g \, := \, [\bby^g_0,     \bby^g_1, ...,           \bby^g_{N-1}] 
          \, := \, [\bbx^g,   \bbS\bbx^g,   ..., \bbS^{N-1}\bbx^g      ] .\!\!
\end{equation}
The matrix $\bbX^g$ is a redundant representation of $\bbx^g$. In fact, for any connected graph any row of $\bbX^g$ is sufficient to recover $\bbx^g$ as each row contains $N$ linear combinations of $\bbx^g$ \cite{marques16-aggregation}. We thus note that any such row has successfully incorporated the graph structure included in the powers of the graph shift operator $\bbS$, without any loss of information. Our goal here is to work at a designated node $p$ with the signal $\bbz_p^g$ that contains the components of the diffusion sequence $\bby^g_k$ that are observed at node $p$ (see second row of Fig.~\ref{fig_aggregation}). This is simply the $p$th row of $\bbX^g$ and leads to the definition
\begin{equation} \label{eqn:agg_representation}
   \bbz_p^g \,:=\ \Big[\bbX^g\Big]_p^{\Tr}
            \, =\ \Big[ \big[          \bbx^g\big]_{p} ; 
                        \big[\bbS      \bbx^g\big]_{p} ; \ldots ; 
                        \big[\bbS^{N-1}\bbx^g\big]_{p} 
                  \Big].
\end{equation}
The signal $\bbz_p^g$ is a local representation at node $p$ that accounts for the topology of the graph in a temporally structured manner. Indeed, since the diffusion sequence $\bby^g_k$ is generated from a temporal diffusion process the components of the sequence $\bbz_p^g$ are elements of a time sequence. Yet, the components of this time sequence depend on the topology of the graph. The first element of $\bbz_p^g$ is the value of the input signal $\bbx^g$ at node $p$, which is independent of the graph topology, but the second element $\bbz_p^g$ aggregates information from values of the input $\bbx^g$ within the neighborhood of $p$ as defined by the nodes that are connected to node $p$. The third element of $\bbz_p^g$ is an aggregate of aggregates which results in the aggregation of information from the 2-hop neighborhood of $p$. As we move forward in the sequence $\bbz_p^g$ we incorporate information from nodes that are farther from $p$ as determined by the topology of the graph. In this way, we have successfully generated a regular structured signal that effectively incorporates the underlying structure. We note that two consecutive elements of $\bbz_{p}^{g}$ indeed relate neighboring values according to the topology of the graph.

%
\begin{figure*}[!t]
\centering
\includegraphics[width=0.2300\textwidth]{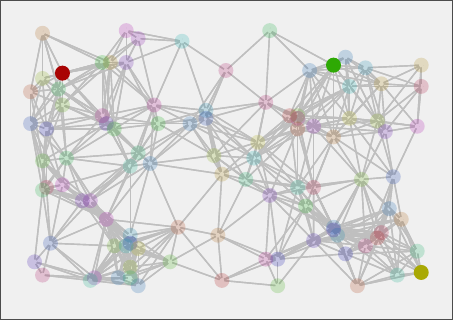} \hfill 
\includegraphics[width=0.2300\textwidth]{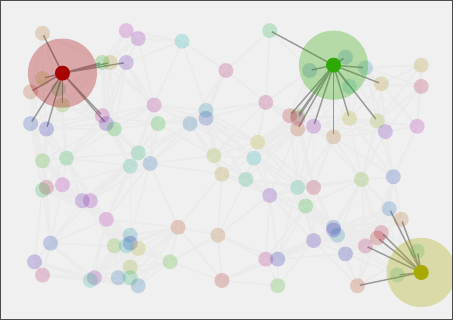} \hfill
\includegraphics[width=0.2300\textwidth]{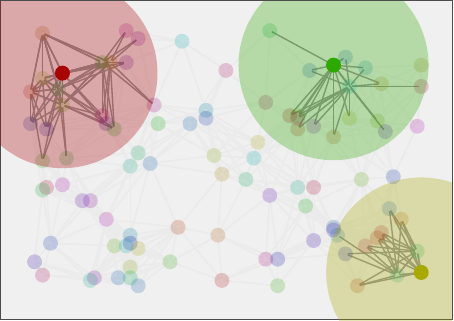} \hfill
\includegraphics[width=0.2300\textwidth]{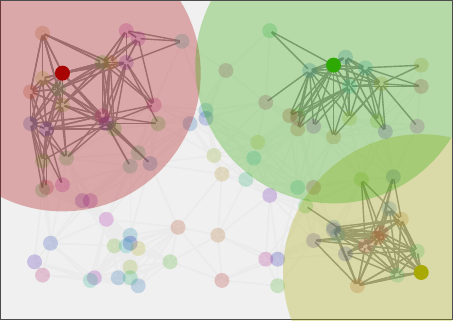} \\ \vspace{2mm}
\includegraphics[width=0.3200\textwidth]{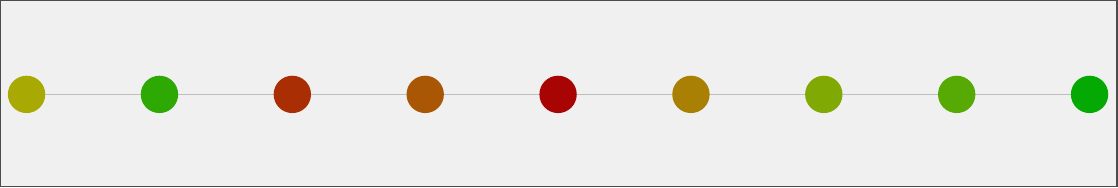}  \hfill
\includegraphics[width=0.3200\textwidth]{figures/MultiNodeCNNGSLayer1Input.pdf}  \hfill
\includegraphics[width=0.3200\textwidth]{figures/MultiNodeCNNGSLayer1Input.pdf}  \\ \vspace{2mm}
\includegraphics[width=0.3200\textwidth]{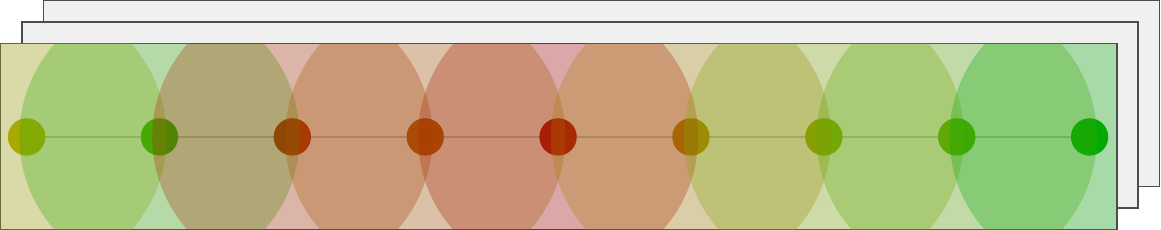}   \hfill
\includegraphics[width=0.3200\textwidth]{figures/MultiNodeCNNGSLayer1Conv.pdf}   \hfill
\includegraphics[width=0.3200\textwidth]{figures/MultiNodeCNNGSLayer1Conv.pdf}   \\ \vspace{2mm}
\includegraphics[width=0.3200\textwidth]{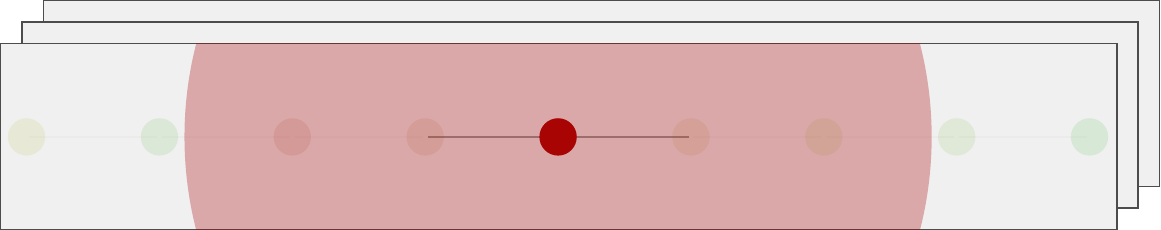}   \hfill
\includegraphics[width=0.3200\textwidth]{figures/MultiNodeCNNGSLayer1Pool.pdf}   \hfill
\includegraphics[width=0.3200\textwidth]{figures/MultiNodeCNNGSLayer1Pool.pdf}   \\ \vspace{2mm}
\includegraphics[width=0.3200\textwidth]{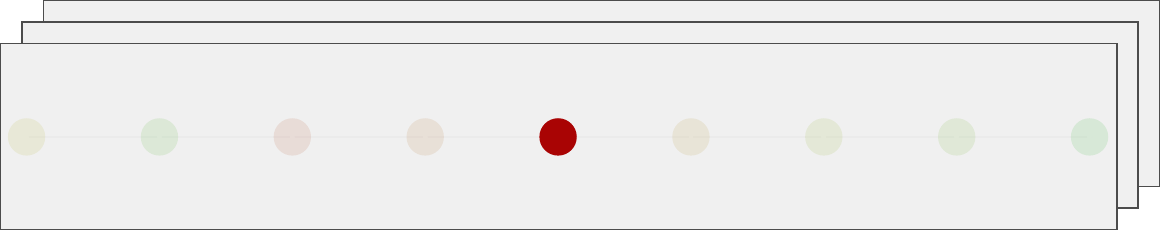}  \hfill
\includegraphics[width=0.3200\textwidth]{figures/MultiNodeCNNGSLayer2Input.pdf}  \hfill
\includegraphics[width=0.3200\textwidth]{figures/MultiNodeCNNGSLayer2Input.pdf}  \\ \vspace{2mm}
\includegraphics[width=0.2300\textwidth]{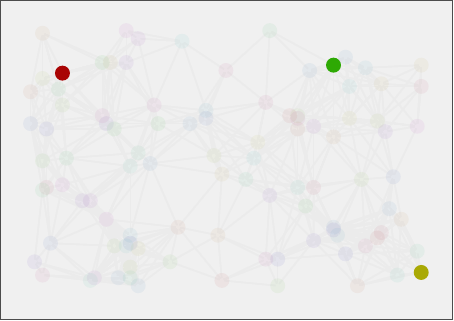} \hfill 
\includegraphics[width=0.2300\textwidth]{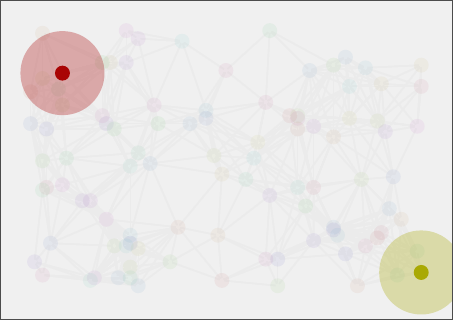} \hfill
\includegraphics[width=0.2300\textwidth]{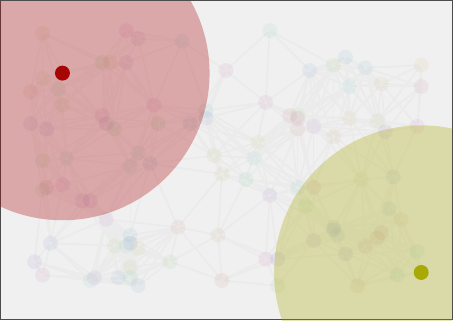} \hfill
\includegraphics[width=0.2300\textwidth]{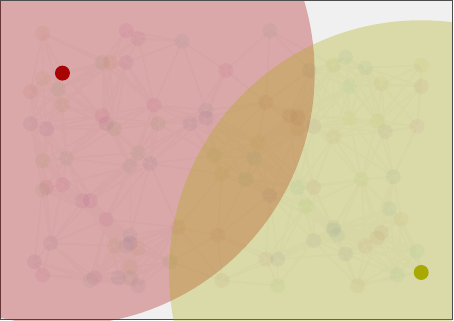} \\ \vspace{2mm}
\caption{Multinode Aggregation Graph Neural Networks. Start by selecting a subset $\ccalP_{1} \subset \ccalV$ of $P_{1}$ nodes of the graph (row 1, diagram 1). Then, proceed to perform $Q_{1}$ local exchanges with neighbors (row 1, diagrams 2, 3, and 4) in order to build $P_{1}$ regular time-structure signals, one at each node (row 2), see \eqref{eqn:multinode_agg_1}. We note that in row 1, the color disks illustrate the reach of the $Q_{1}$ local exchanges of each of the selected nodes $\ccalP_{1}$. Once the regular structured signals have been constructed on each of the $P_{1}$ nodes, proceed with a regular CNN, applying regular convolution (row 3), and regular pooling (row 4), until $F_{L_{1}}$ features are obtained at each node (row 5), see \eqref{eqn:conv_time}-\eqref{eqn_downsampling}, \eqref{eqn:inner_layer_output}. Now, we view each feature as a graph signal being supported on the selected nodes, see \eqref{eqn:outer_layer_input}, zero-padded to fit the graph (row 6, diagram 1), see \eqref{eqn:outer_layer_pad}. We then select a smaller subset $\ccalP_{2} \subseteq \ccalP_{1}$ of $P_{2} \leq P_{1}$ nodes (row 2, diagram 2) and carry out $Q_{2}$ local exchanges with the neighbors, (row 2, diagrams 2, 3 and 4), illustrated with color disks in the last row. These neighbor exchanges create new regular structured signals at each of the $\ccalP_{2}$ nodes, see cf.~\eqref{eqn:outer_layer_time}. Then, we continue by computing $F_{L_{2}}$ regional features at each node by means of regular CNNs and so on.}
	\label{fig_multinode}
\end{figure*}

If the signal $\bbz_p^g$ is a signal in time, it can be processed with a regular CNN architecture and this is indeed our definition of aggregation GNNs. At the first layer $\ell=1$ we take the locally aggregated signal $\bbz_p^g$ as input and produce features $\bbu_{p1}^{fg}$ by convolving with the $K_{p1}$-tap filter $\bbh_{p1}^{fg}$ [cf. \eqref{eqn:conv_time}],
\begin{align}\label{eqn_conv_agg_layer_1}
   \Big[\bbu_{p1}^{fg}\Big]_n 
       :=   \Big[\bbh_{p1}^{fg} \ast \bbz_p^g\Big]_{n}  
	    =   \sum_{k=0}^{K_{p1}-1}  \Big[ \bbh_{p1}^{fg}  \Big]_{k}  \,
	        \Big[ \bbz_p^g \Big]_{n-k} ,
\end{align}
where we use zero padding to account for border effects and assume the size of the output is the same as the input. The convolution in \eqref{eqn_conv_agg_layer_1} is the regular time convolution. In fact, except for minor notational differences to identify the aggregation node $p$, \eqref{eqn_conv_agg_layer_1} is the same as \eqref{eqn:conv_time} with $\ell=1$. The topology of the graph is incorporated in \eqref{eqn_conv_agg_layer_1} not because of the convolution but because of the way in which we construct $\bbz_p^g$. To emphasize the effect of the topology of the graph we use \eqref{eqn:agg_representation} to rewrite \eqref{eqn_conv_agg_layer_1} as
\begin{equation} \label{eqn_conv_agg_layer_1_shift_explicit}
   \Big[\bbu_{p1}^{fg}\Big]_n 
       = \sum_{k=0}^{K_{p1}-1} 
              \Big[ \bbh_{p1}^{fg}     \Big]_{k} 
              \Big[ \bbS^{n-k-1}\bbx^g \Big]_{p}
\end{equation}
Since the convolution in \eqref{eqn_conv_agg_layer_1_shift_explicit} considers consecutive values of the signal $\bbz_p^{g}$, the features $\bbu_{p1}^{fg}$ have a structure that is convolutional on the graph $\bbS$. Each feature element $[\bbu_{p1}^{fg}]_{n}$ is a linear combination of consecutive $K_{p1}$ neighboring values of the input $\bbx^{g}$ starting with shift $\bbS^{n-1}\bbx^{g}$ and ending at $\bbS^{n-K_{p1}-1}\bbx^{g}$. Alternatively, note that the regular convolution operation linearly relates consecutive elements of a vector; and since consecutive elements in vector $\bbz_{p}^{g}$ reflect nearby neighborhoods according to the graph, we have effectively related neighboring values of the graph signal by means of a regular convolution. Thus, coefficients $\bbh_{p1}^{fg}$ encoding low-pass filters further aggregate information across neighborhoods, while high-pass filters output features quantifying differences between consecutive neighborhood resolutions. Thus, low-pass time filters applied to $\bbz_p^{g}$ detect features that are smooth on the graph $\bbS$, while high-pass time filters applied to $\bbz_p^{g}$ detect sharp transitions between signal values between nearby nodes.

Once the features $\bbu_{p1}^{fg}$ in \eqref{eqn_conv_agg_layer_1}, or their equivalents in \eqref{eqn_conv_agg_layer_1_shift_explicit}, are computed, we sum features $\bbu_{p1}^{fg}$ as per \eqref{eqn:agg_features} obtaining $\bbu_{p1}^{f}$, compute local summaries as per \eqref{eqn_group_nonlinearity} yielding $\bbv_{p1}^{f}$, and subsample according to \eqref{eqn_downsampling} resulting in features $\bbx_{p1}^{f}$. Since in this case the indexes of the feature vector represent (neighborhood) resolution, some applications may benefit from non-equally spaced sampling schemes that put more emphasis on sampling the high-resolution (low-resolution) part of the feature vector. Subsequent layers repeat the computation of convolutional features and pooling steps in \eqref{eqn:conv_time}-\eqref{eqn_downsampling}. Formally, all of the variables in \eqref{eqn:conv_time}-\eqref{eqn_downsampling} need to be marked with a subindex $p$ to identify the aggregation node.

\begin{remark} \normalfont
The aggregation GNN architecture reduces trivially to conventional CNNs when particularized to graph signals defined over a cyclic graph. Since $[\bbA_{\dc}^{k} \bbx^{g}]_{p} = [\bbx^{g}]_{1+(p+k) \mod N}$ is a cyclic shift of the input signal $\bbx^{g}$, then $\bbz_{p}^{g}=\bbx^{g}$ in \eqref{eqn:agg_representation} for all $p$ and a regular CNN follows.
\end{remark}

\begin{remark} \normalfont
The aggregation GNN architecture rests on transforming the data on the graph in such a way that it becomes supported on a regular structure, and thus regular CNN techniques can be applied. Transforming graph data is the main concern of graph embeddings \cite{cai17-embeddings}. Unlike the methods surveyed in \cite{cai17-embeddings}, we consider the underlying graph support $\ccalG$ as given (not learned), we do not attempt to compress the graph data as construction of aggregated vector $\bbz_{p}^{g}$ does not entail any loss of information (if all eigenvalues of $\bbS$ are distinct), and the focus is on data defined on top of the graph (the graph signal), rather than the graph itself (given by $\bbS$).
\end{remark}

%
\subsection{Multinode Aggregation Graph Neural Networks}\label{sec_aggregation_multinode}

Selecting a single node $p \in \ccalV$ to aggregate all the information generally entails $N-1$ local exchanges with neighbors [cf. \eqref{eqn:aggregation_matrix}]. For large networks, carrying out all these exchanges might be infeasible, either due to the associated communication overhead or due to numerical instabilities. This can be overcome by selecting a subset of nodes to aggregate local information, i.e., selecting a submatrix of \eqref{eqn:aggregation_matrix} with a few rows and columns in lieu of a single row with all the columns; see Fig.~\ref{fig_multinode}. The selected nodes will first process their own samples using an aggregation GNN and then exchange the obtained outputs with the other selected nodes. This process is repeated until the information has been propagated through the entire graph. 

To explain such a two-level hierarchical architecture, let us denote as $\ell$ the layer index for the aggregation stage and as $r$ the layer index for the exchange stage. The total number of exchange (outer) layers is $R$ and, for each outer layer $r$, a total number of $L_r$ aggregation (inner) layers is run. We start by describing the procedure for $r=1$, where $\ccalP_{1} \subset \ccalV$ denotes the subset of selected nodes and let $Q_{1}$ denotes the number of times the shift is applied ($\bbS^{q}$, for $q=0,\ldots,Q_{1}-1$). It is observed that this amounts to selecting $P_1=|\ccalP_{1}|$ rows and $Q_{1}$ consecutive columns of \eqref{eqn:aggregation_matrix}. Setting $\ell=0$, the signal aggregating the $(Q_{1}-1)$-hop neighborhood information at each node $p \in \ccalP_{1}$ can be constructed as [cf.~\eqref{eqn:agg_representation}]
\begin{equation} \label{eqn:multinode_agg_1}
	\bbz_{p0}^{g}(1,Q_{1}) 
			\,:=\ \Big[ \big[          \bbx^g\big]_{p} ; 
                        \big[\bbS      \bbx^g\big]_{p} ; \ldots ; 
                        \big[\bbS^{Q_{1}-1}\bbx^g\big]_{p} 
                  \Big].
\end{equation}
Since $\bbz_{p0}^{g}$ exhibits a time structure, the regular CNN steps \eqref{eqn:conv_time}-\eqref{eqn_downsampling} can be applied individually at each node (see Fig.~\ref{fig_multinode}). More specifically, since $L_1$ denotes the number of layers for the aggregation stage when $r=1$, a set of $F_{L_{1}}$ descriptive features of the $(Q_{1}-1)$-hop neighborhood of node $p$ is constructed by concatenating $\ell=0,\ldots,L_{1}-1$ layers of the form \eqref{eqn:conv_time}-\eqref{eqn_downsampling} as is done in the aggregation GNN. Setting $\ell=L_1$, the output of the last layer of the aggregation stage is
\begin{equation} \label{eqn:inner_layer_output}
	\bbz_{pL_{1}}(1,Q_{1}) 
		= \Big[ z_{pL_{1}}^{0}; \ldots; z_{pL_{1}}^{F_{L_{1}}} \Big].
\end{equation}
Different feature vectors $\bbz_{p L_{1}}$ of dimension $F_{L_{1}}$ are obtained at each of the $p$ selected nodes, describing the corresponding $(Q_{1}-1)$-hop neighborhood.

In order to further aggregate these local features (describing local neighborhoods) into more global information, we need to collect each feature $g$ at every selected node $p \in \ccalP_{1}$. More precisely, let $P_{1} = |\ccalP_{1}|$ be the number of selected nodes, then
\begin{equation} \label{eqn:outer_layer_input}
	\bbx_{1}^{g} = \Big[ z_{p_{1}L_{1}}^{g};\ldots;z_{p_{P_{1}}L_{1}}^{g} \Big]
\end{equation}
where each $p_{k} \in \ccalP_{1}$. We now set $r=2$ and select a subset of nodes $\ccalP_{2} \subseteq \ccalP_{1}$ to aggregate features $\bbx_{1}^{g}$ by means of local neighborhood exchanges. However, signal $\bbx_{1}^{g}$ has dimension $P_{1} < N$, so it cannot be directly exchanged through the original graph $\ccalG$. We therefore use zero padding to make $\bbx_{1}^{g}$ fit the graph
\begin{equation} \label{eqn:outer_layer_pad}
	\tbx_{1}^{g} = \bbP_{1}^{\Tr} \bbx_{1}^{g}
\end{equation}
with $\bbP_{1} $ being the ${P_{1} \times N}$ fat binary matrix that selects the subset $\ccalP_{1}$ of rows of \eqref{eqn:aggregation_matrix}. Then, we apply $Q_{2}$ times the original shift $\bbS$ to the signals $\tbx_{1}^{g}$, collecting information at nodes $p \in \ccalP_{2}$,
\begin{equation} \label{eqn:outer_layer_time}
	\bbz_{p0}^{g}(2,Q_{2}) 
			\,:=\ \Big[ \big[          \tbx_{1}^g\big]_{p} , 
                        \big[\bbS      \tbx_{1}^g\big]_{p} , \ldots , 
                        \big[\bbS^{Q_{2}-1}\tbx_{1}^g\big]_{p} 
                  \Big]^{\Tr}.
\end{equation}
Once $\bbz_{p0}^{g}$ is collected at each node $p \in \ccalP_{2}$ the time-structure of the signal is exploited to deploy another regular CNN \eqref{eqn:conv_time}-\eqref{eqn_downsampling} (aggregation GNN stage) in order to obtain $F_{L_{2}}$ features that describe the region.

In general, consider the output of \emph{outer layer} $r-1$ is $\bbx_{r-1}^{g}$, consisting of feature $g$ at a subset $\ccalP_{r-1}$ of $P_{r-1}$ nodes [cf. \eqref{eqn:outer_layer_input}], for $g=1,\ldots,F_{L_{r-1}}$. Then, this signal is zero padded to fit the original graph $\tbx_{r-1}^{g}=\bbP_{r-1}^{\Tr}\bbx_{r-1}^{g}$ [cf.~\eqref{eqn:outer_layer_pad}] and the graph shift $\bbS$ is applied $Q_{r}$ times, collecting the shifted versions at a subset of nodes $\ccalP_{r}$ to construct time-structure signal $\bbz_{p0}^{g}(r,Q_{r})$ [cf. \eqref{eqn:outer_layer_time}]. Each node $p \in \ccalP_{r}$ runs a regular CNN \eqref{eqn:conv_time}-\eqref{eqn_downsampling} with $L_{r}$ \emph{inner layers} to produce $F_{L_{r}}$ features $\bbz_{pL_{r}}(r,Q_{r})$ [cf. \eqref{eqn:inner_layer_output}] that are then collected at each of the nodes $p \in \ccalP_{r}$ to produce $\bbx_{r}^{f}$ [cf.~\eqref{eqn:outer_layer_input}], for $f=1,\ldots,F_{L_{r}}$. See Fig.~\ref{fig_multinode} for an illustration of the architecture.

%
\subsection{Practical Considerations}

\myparagraph{Local architecture.} The single node aggregation GNN architecture is entirely \emph{local}. Only one node $p \in \ccalV$ is selected, and that node gathers all the relevant information about the data by means of local exchanges only. Furthermore, the output at the last layer is also obtained at a single node, so there is no need to have actual physical access to every node in the network.

\myparagraph{Regular CNN design.} Since signal $\bbz_{p}^{g}$ gathered at node $p$ exhibits a regular time structure, the state-of-the-art expertise in designing conventional CNNs can be immediately leveraged to inform the design of convolutional layers of aggregation GNNs.

\myparagraph{Numerical normalization.} For big networks, some of the entries of $\bbS^{k}$ (as well as the components of $\bbz_p^{g}$ associated with those powers) can grow too large, leading to numerical instability. To avoid this, aggregation schemes typically work with a normalized version of the graph shift operator that guarantees that the spectral radius of $\bbS$ is one. 
  
\myparagraph{Choice of aggregating node.} The choice of nodes that aggregate all the information has an impact on the overall performance of the algorithm. This decision can be informed by several criteria such as the degree, the frequency content of the signals of interest \cite{marques16-aggregation} or be determined by different measures of centrality in the network \cite{segarra16-centrality}. In particular, in the experiments carried out in Sec.~\ref{sec:sims}, we select nodes based on the leverage scores obtained by the two sampling schemes described in \cite{anis16-spectralproxies} and \cite{varma15-scores}.

\myparagraph{Filter taps.} For a generic (inner) layer $1<\ell<L_{r}$ the generation of the feature vectors $\bbu_{\ell}^{fg}\in \reals^{N_{\ell-1}}$ and $\bbu_{\ell}^{f}\in \reals^{N_{\ell-1}}$ is as in \eqref{eqn:conv_time} and \eqref{eqn:agg_features}, so that we have that $\bbu_{\ell}^{f}\ =\ \sum_{g=1}^{F_{\ell-1}} \bbu_{\ell}^{fg}\  =\ \sum_{g=1}^{F_{\ell-1}} \bbh_{\ell}^{fg} \ast \bbz_{p(\ell-1)}^{g}$. The main difference in this case is on the type and length of the filter coefficients $\bbh_{\ell}^{fg}\in \reals^{K_{\ell}}$. While in classical CNNs the filter coefficients are critical to aggregate the information at different resolutions, here part of that aggregation has been already taken care of in the first layer when transforming $\bbx^{g}$ into $\bbz_p^{g}$. As a result, the filter taps in the aggregation GNN architecture can have a shorter length and place more emphasis in high frequency features.

\myparagraph{Pooling.} Something similar applies to the pooling schemes. The summarization and downsampled vectors for the aggregation architecture are obtained as $[\bbv^{f}_{\ell}]_n 
= \rho_{\ell}( [\bbu_{\ell}^f]_{\bbn_\ell})$ and $\bbx^{f}_{\ell} =  \sigma_\ell (\bbC_{\ell}\bbv^{f}_{\ell} )$, which coincide with their counterparts for classical CNNs in \eqref{eqn_group_nonlinearity} and \eqref{eqn_downsampling}. The difference is therefore not in the expressions, but on how $\bbn_{\ell}$ and $\bbC_{\ell}$ are selected. 
While in traditional CNNs the signal $\bbx^{g}$ is global in that all the samples have the same resolution, in the aggregation architecture the signal $\bbz_p^{g}$ is local and different samples correspond to different levels of resolution. More specifically, aggregation pooling schemes for $\bbn_{\ell}$ and $\bbC_{\ell}$ that preserve the top samples of the feature vectors $\bbu^f_\ell$ to keep finer resolutions combined with a few bottom samples to account for coarser information are reasonable, while in traditional CNNs regular schemes for $\bbn_{\ell}$ and $\bbC_{\ell}$ that extract information and sample the signal support regularly can be more adequate.

\myparagraph{Design flexibility.} The multinode aggregation GNN acts as a decentralized method for constructing regional features. We note that, for ease of exposition, the number of shifts $Q_{r}$ at each outer layer is the same for all nodes as well as the number of features $F_{L_{r}}$ that are obtained at each node. However, this architecture can be adapted to different node-dependent parameters in a straightforward manner.

\myparagraph{Computational cost.} The computational cost of the multinode aggregation GNN at each outer layer $r$ is that of processing the regular CNN for each node, $O (\sum_{p=1}^{P_{r}} \sum_{\ell=1}^{L_{r}} N_{\ell-1} K_{\ell} F_{\ell-1} F_{\ell})$ which can be easily distributed among the $P_{r}$ involved nodes.

\myparagraph{Number of parameters.} The number of parameters of the multinode aggregation GNN is $O(\sum_{p=1}^{P_{r}} \sum_{\ell=1}^{L_{r}} K_{\ell} F_{\ell} F_{\ell-1})$.  We observe, though, that the regular CNNs employed tend to be very small, since the initial $Q_{r}$ regular CNN at each node) as well as the length of the filters $K_{\ell}$ are very small as well (typically, $K_{\ell} \ll Q_{r}$, cf. Sec.~\ref{sec:regular}).


\section{Numerical Experiments} \label{sec:sims}



We test the proposed GNN architectures and compare their performance with the graph coarsening (multiscale hierarchical clustering) approach of \cite{defferrard17-cnngraphs}. In the first scenario (Sec.~\ref{subsec_sourceloc}), we address the problem of source localization on synthetic stochastic block model (SBM) networks. Then, we move the source localization problem to a more realistic setting of a Facebook network of $234$ users (Sec.~\ref{subsec_fb}). As a third experiment, we address the problem of authorship attribution (Sec.~\ref{subsec_author}). And finally, we test the proposed architectures in the problem of text categorization on the \texttt{20NEWS} dataset (Sec.~\ref{subsec_20news}).

We test the proposed Selection (Sec.~\ref{sec:selection}), Aggregation (Sec.~\ref{sec:aggregation}) and Multinode (Sec.~\ref{sec_aggregation_multinode}) GNN architectures. For the selection of nodes involved in each of the architectures, we test three different strategies. First, we choose nodes based on their degree; second, we select them following the leverage scores proposed by the experimentally designed sampling (EDS) in \cite{varma15-scores}; and third, we determine the appropriate nodes by using the spectral-proxies approach (SP) in \cite{anis16-spectralproxies}.  In all architectures, the last layer is a fully-connected readout layer, followed by a softmax, to perform classification.

Unless otherwise specified, all GNNs were trained using the ADAM optimizer \cite{kingma17-adam} with learning rate $0.001$ and forgetting factors $\beta_{1}=0.9$ and $\beta_{2}=0.999$. The training phase is carried out for $40$ epochs with batches of $100$ training samples. The loss function considered in all cases is the cross-entropy loss between one-hot target vectors and the output from the last layer of each architecture, interpreted as probabilities of belonging to each class. Also, in all cases, we consider max-pooling summarizing functions and ReLU activation functions for the corresponding GNN layers.

\begin{table}
	\centering
\begin{tabular}{lc} \hline
Architecture 				& Accuracy 				\\ \hline
Selection (S) Degree		& $86.9 (\pm 5.9) \%$	\\
Selection (S) EDS			& $90.0 (\pm 4.6) \%$	\\
Selection (S) SP			& $91.1 (\pm 4.7) \%$	\\
Aggregation	(A)	Degree		& $94.2 (\pm 4.7) \%$	\\
Aggregation	(A)	EDS			& $96.5 (\pm 3.1) \%$	\\
Aggregation	(A)	SP			& $95.2 (\pm 4.4) \%$	\\
Multinode (MN) Degree		& $96.1 (\pm 3.4) \%$	\\
Multinode (MN) EDS			& $96.0 (\pm 3.5) \%$	\\
\textbf{Multinode (MN) SP}			& $\mathbf{97.3 (\pm 2.7) \%}$	\\
Graph Coarsening (C) Clustering		& $87.4 (\pm 3.2)\%$ \\ \hline
\end{tabular}
	\caption{Considering that SBM graphs are random, we generate $10$ different instances of SBM graphs with $N=100$ nodes and $C=5$ communities of $20$ nodes each. For each of these $10$ graphs, we randomly generate $10$ different datasets (training, validation and test set). We compute the classification accuracy of each realization, and average across all $10$ realizations for each graph, obtaining $10$ average classification accuracies. In the table we show the classification accuracy, averaged across the $10$ graph instances. The standard deviation from these $10$ graphs is also shown.}
	\label{table_sourceloc}
\end{table}


\subsection{Source Localization} \label{subsec_sourceloc}

Consider a connected stochastic block model (SBM) network with $N=100$ nodes and $C=5$ communities of $20$ nodes each \cite{decelle11-sbm}. In SBM graphs, edges are randomly drawn between nodes within the same community, independently, with probability $0.8$; while edges are randomly drawn between nodes of different communities, independently, with probability $0.2$. Denote by $\bbA$ the adjacency matrix of such graph.

\begin{figure*}
\centering
\begin{subfigure}{.33\textwidth}
  \centering
  \includegraphics[width=0.95\textwidth]{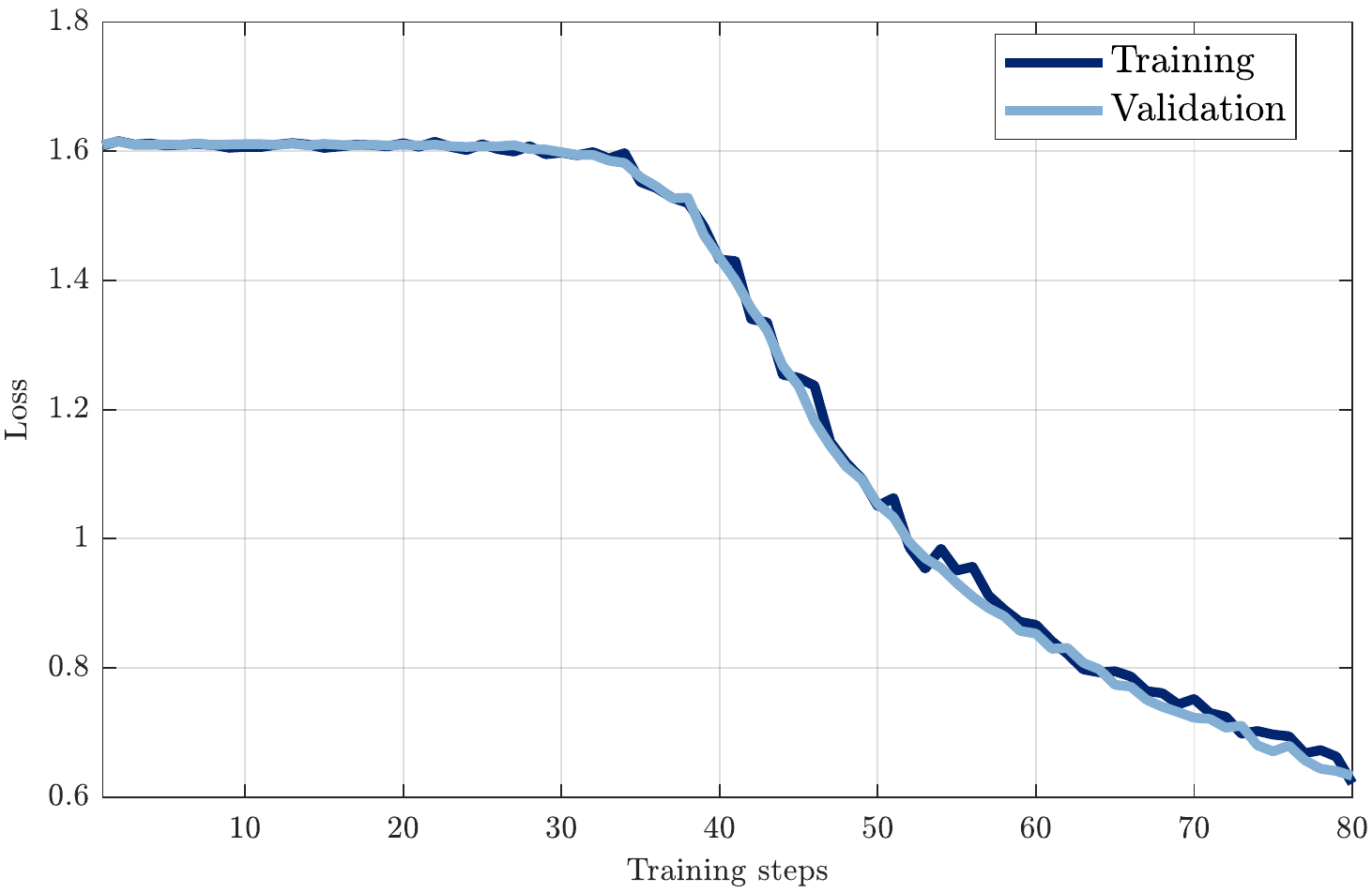}
  \caption{Selection GNN SP}
  \label{fig:a}
\end{subfigure}%
\hfill
\begin{subfigure}{.33\textwidth}
  \centering
  \includegraphics[width=0.95\textwidth]{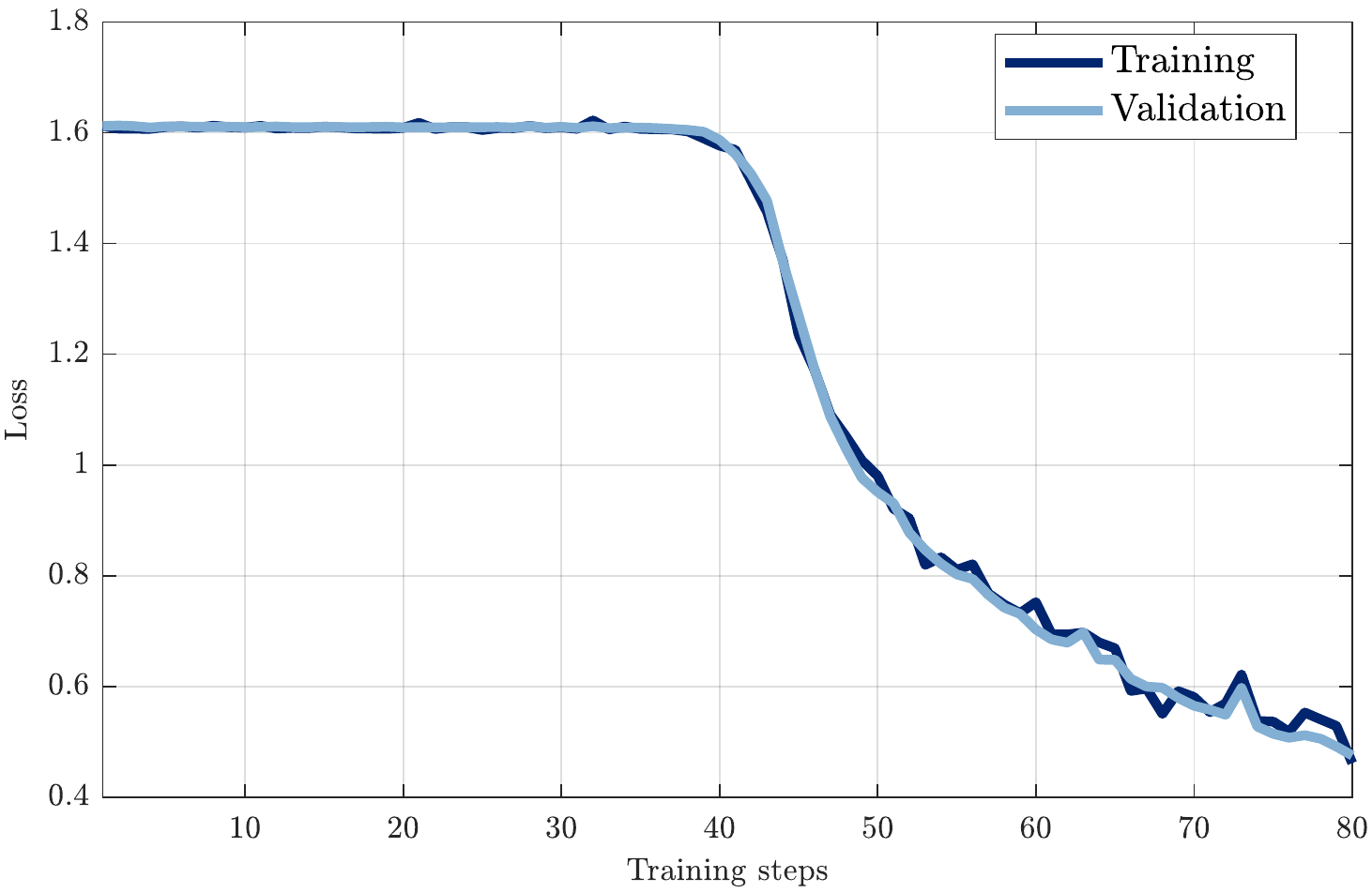}
  \caption{Aggregation GNN EDS}
  \label{fig:mn_P}
\end{subfigure}%
\hfill
\begin{subfigure}{.33\textwidth}
  \centering
  \includegraphics[width=0.95\textwidth]{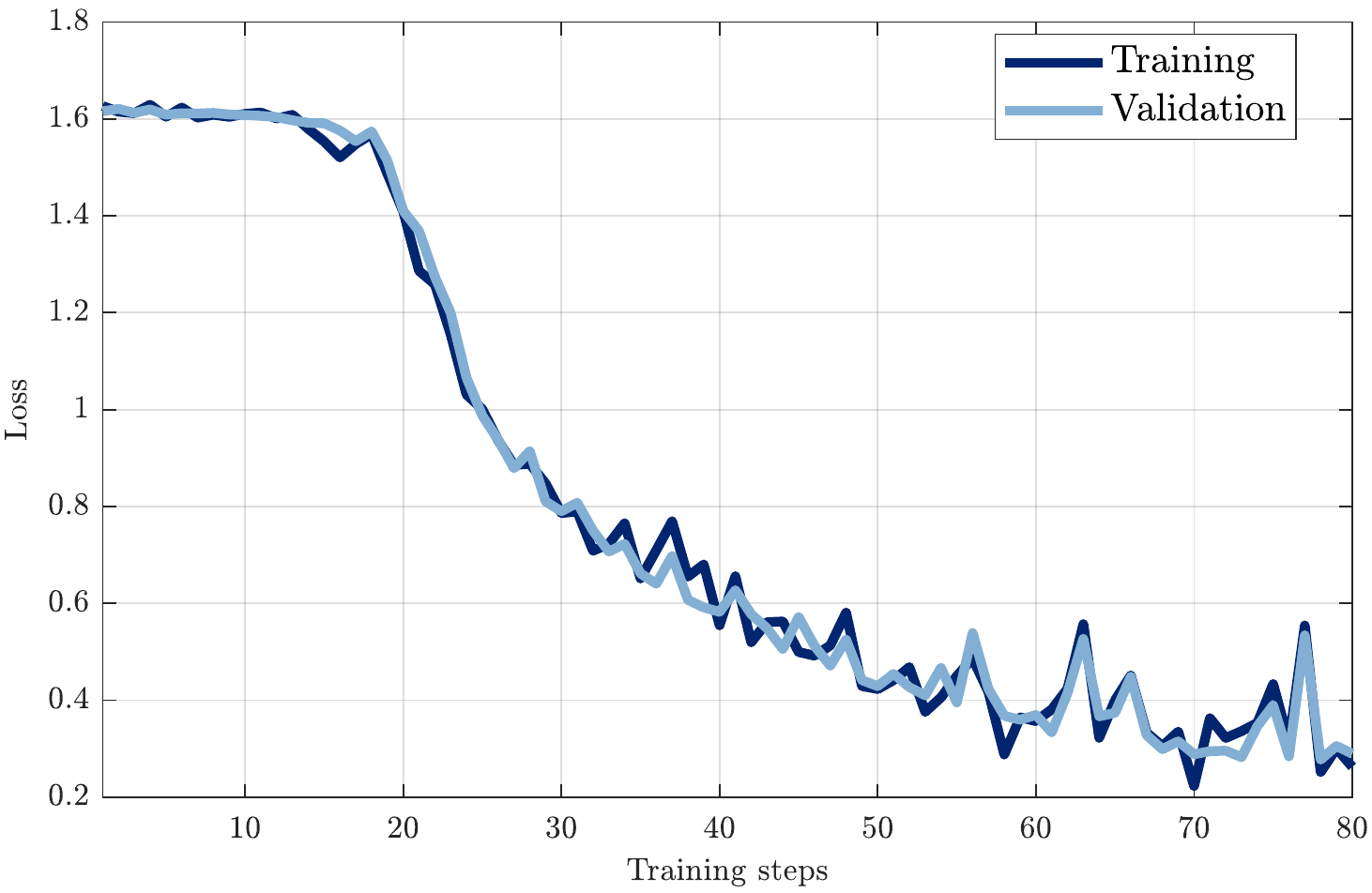}
  \caption{Multinode GNN SP}
  \label{fig:mn_Q}
\end{subfigure}%
\caption{Validation and training loss during training stage. We observe that the validation loss and the training loss are essentially equal throughout the training stage for all three architectures. This shows that the proposed models are not overfitting the data, since the validation loss keeps decreasing with the training steps The best performing selection method of each architecture is represented.}
\label{fig:validation}
\end{figure*}

In the problem of source localization, we observe a signal that has been diffused over the graph and estimate the spatial origin of such diffused process. More precisely, consider $\bbdelta_{c}$ a graph signal that has a $1$ at node $c$ and $0$ at every other node. Define $\bbx = \bbA^{t} \bbdelta_{c}$ as the diffused graph signal, for some unknown $t \geq 0$. The objective is to estimate the origin $c$ of the diffusion. In this situation in particular, we are interested in estimating the \emph{community} $c$ (rather than the node) that originated the observed signal $\bbx$. We can thus model this scenario as a classification problem in which we observe graph signal $\bbx$ and have to assign it to one of the $C=5$ communities.

In the simulations, we generate the training and test set by randomly selecting the origin $c$ from a pool of $C=5$ nodes (the largest-degree node of each community; recall that all nodes have, on average, the same degree) and randomly selecting the diffusion time $t<25$, as well. We generate a training set of $10,000$ signals and a test set of $200$ signals. The training set is further split in $2,000$ signals for validation, and the rest for training. We simulate $10$ graphs, and for each graph, we simulate $10$ realizations of training and test sets. For numerical reasons, the adopted graph shift operator is $\bbS = \bbA/\lambda_{\max}$ where $\lambda_{\max}$ is the maximum eigenvalue of $\bbA$.

The architectures tested are as follows. For the selection GNN we consider two layers selecting $10$ nodes in each. The number of output features in each layer are $F_{1} = F_{2} = 32$ and the filters consists of $K_{1} = K_{2} = 5$ taps [cf. \eqref{eqn_conv_features_unpadded}]. For the summarizing functions, we consider neighborhoods of size $\alpha_{1} = 6$ and $\alpha_{2} = 8$, respectively [cf. \eqref{eqn_graph_neigborhood}]. In the aggregation GNN, we select the single node with highest: a) degree, b) EDS leverage score, or d) spectral-proxies (SP) norm, depending on the strategy chosen. Then, we construct the regular-structured signal [cf. \eqref{eqn:agg_representation}] and apply the aggregation GNN with two layers. The number of features on each layer is $F_{1}=16$ and $F_{2} = 32$, with filters of size $K_{1} = 4$ and $K_{2}=8$ [cf. \eqref{eqn_conv_agg_layer_1_shift_explicit}]. Max-pooling is applied to reduce the size of the regular signal by half on each layer, and the nonlinearity used is the ReLU. Finally, for the multinode GNN, we consider two outer layers selecting $P_{1}=10$ and $P_{2}=5$ nodes and shifting the signal $Q_{1}=7$ and $Q_{2}=5$ times to build the regular signal on each node [cf. \eqref{eqn:multinode_agg_1}]. Then, for each outer layer, we apply two inner layers. In the first one, we obtain $16$ features at each inner layer; and in the second outer layer, we get $16$ and $32$ for each inner layer. In all inner layers, the filters are of size $3$, with max-pooling by $2$, and a ReLU nonlinearity. We recall that the selection of nodes depends on the sampling strategy selected (degree, EDS or SP). We compare against a two-layer architecture using graph coarsening \cite{defferrard17-cnngraphs}, reducing the number of nodes to a half on each layer, computing $F_{1}=F_{2}=32$ features with filters consisting of $K_{1}=K_{2}=5$ filter taps. In contrast with the previous cases where $\bbS$ was set to the rescaled adjancency matrix, in the graph coarsening architecture we set $\bbS$ to normalized Laplacian, since that was the specification in \cite{defferrard17-cnngraphs} and, more importantly, yields a better performance.

The plots in Fig.~\ref{fig:validation} show the value of the loss function on the training and validation sets as the training stage progresses. We observe that both drop with training, showing that the model is effectively learning from data. We see that it takes some time for the models to start learning (reaching half of the training stage in the case of aggregation), but then effectively lower the training loss. We also see that the Multinode GNN achieves a lower loss value, which translates in better performance on the test set, and that it also takes the least number of training steps before starting to lower the loss function. Finally, we note that the validation loss and the training loss are essentially the same, showing that there is no overfit in the models.

Accuracy results on the test set are presented in Table \ref{table_sourceloc}. The accuracy results for all $10$ realizations of each graph are averaged, and then all $10$ graph mean accuracies are averaged to obtain the values shown in Table \ref{table_sourceloc}. The error values in the table are the square root of the variance computed across the means obtained for each of the $10$ graphs. We observe that the best performance is achieved by Multinode GNN with nodes chosen following the spectral proxies method. We observe that all multinode and aggregation GNNs outperform the graph coarsening approach, and so do selection GNNs following EDS and spectral proxies sampling.


\subsection{Facebook network} \label{subsec_fb}

For this second experiment, we also consider the source localization problem, but in this case, we test it on top of a real-world network. In particular, we built a 234-user Facebook network as the largest connected network within the larger dataset provided in \cite{mcauley12-fb}. We observe that the resulting network exhibits two communities of quite different size. The source localization problem formulation is the same than the one described in the previous section, where the objective is to identify which of the two communities originated the diffusion process. This is analogous to finding the start of a rumor. Again, we set $\bbS = \bbA/\lambda_{\max}$. The datasets are generated in the same fashion as described in the previous section.

The three architectures used are as follows. For the selection GNN we use two layers, choosing $10$ nodes after the first one, and use filters with $K_{1}=K_{2}=5$ taps that generate $F_{1}=F_{2}=32$ features on each layer. For the pooling stage, we use a $\max\{\cdot\}$ summarization with $\alpha_{1}=2$ and $\alpha_{2}=4$. In the aggregation GNN we select the best node based on one of the three sampling strategies (degree, EDS and SP) and the gather the regular-structure data at that node. We then process it with a two-layer CNN that generates $F_{1}=32$ and $F_{2}=64$ features, using $K_{1}=K_{2}=4$. Max-pooling of size $2$ is used on each layer (i.e. half of the samples gathered at the node are kept after each layer). In the case of the multinode GNN we use two-outer layers, selecting $P_{1}=30$ and $P_{2}=10$ nodes on each, and gathering $Q_{1}=Q_{2}=5$ shifted versions of the signal at each node. Then, for the inner layers, we use two-layer architectures that generate $16$ features on each layer in the first outer layer, and $16$ and $32$ features on each layer in the second outer layer. In all cases, we use filters of size $3$ and max-pooling by a factor of $2$. Finally, for the graph coarsening architecture, we adopt the normalized Laplacian as GSO, as described in \cite{defferrard17-cnngraphs}, and use two-layers computing $F_{1}=F_{2}=32$ features using graph filters with $K_{1}=K_{2}=5$ filter taps. After each layer, the number of nodes is reduced by half.

\begin{table}
	\centering
\begin{tabular}{lc} \hline
Architecture 				& Accuracy 				\\ \hline
Selection (S) Degree		& $96.0 (\pm 1.5) \%$	\\
Selection (S) EDS			& $95.6 (\pm 1.0) \%$	\\
Selection (S) SP			& $97.6 (\pm 1.2) \%$	\\
Aggregation	(A)	Degree		& $95.8 (\pm 1.6) \%$	\\
Aggregation	(A)	EDS			& $96.9 (\pm 1.2) \%$	\\
Aggregation	(A)	SP			& $95.8 (\pm 1.4) \%$	\\
Multinode (MN) Degree		& $97.6 (\pm 1.3) \%$	\\
Multinode (MN) EDS			& $96.8 (\pm 1.2) \%$	\\
\textbf{Multinode (MN) SP}			& $\mathbf{99.0 (\pm 0.8) \%}$	\\
Graph Coarsening (C) Clustering		& $95.2 (\pm 1.2)\%$ \\ \hline
\end{tabular}
	\caption{Classification accuracy averaged across $10$ different realizations of the training and test sets for the same underlying graph. In parenthesis, we show the standard deviation of the classification accuracy.}
	\label{table_fb}
\end{table}

For training we use $80$ epochs. We also generate $10$ different random realizations of the dataset to account for random variabilities in the setting. Results for all ten architectures are shown in Table \ref{table_fb}. We observe that all architectures achieve a very high classification accuracy. We note that selection GNN tends to outperform aggregation GNN. The best result is obtained for multinode GNN using spectral proxies and is $99.0\%$ classification accuracy.


\subsection{Authorship attribution} \label{subsec_author}

As a third experiment, we study the problem of authorship attribution, as detailed in \cite{segarra15-wans}. We consider excerpts of works written by a myriad of contemporary authors from the 19th century. We then build a word adjacency network (WAN) using functional words in these excerpts, and obtain a graph profile for each author, i.e., a graph that represents an author's writing style by the way functional words (who act as nodes) are linked (weighted edges) in the excerpts written; see \cite{segarra15-wans} for a full detail on the authors considered and the specific construction of WANs. Then, we take a new excerpt, of unknown authorship, and by looking at the frequency of the functional words, we want to determine who the author is. In the framework presented in this paper, the signature word adjacency network constitutes the underlying graph support, and the frequency count of functional words becomes the graph signal.

In particular, we focus on texts authored by Emily Bront{\"{e}}. We consider a corpus of $682$ excerpts of around $1000$ words, authored by her; and take into consideration $211$ functional words. Then, we take $546$ of these excerpts as a training set, in order to both, build the signature WAN, and also as training samples. The constructed graph consists of $N=211$ nodes, one for each functional word, the edges and their weights are determined by the precedence relationship between each word, as described in \cite{segarra15-wans}; and each training sample consist of a graph signal, where the value associated to each node is the frequency count of that specific functional word. The remaining $136$ excerpts are used as test samples. Once the signature WAN for Bront\"{e} is built, we construct a training set of $1092$ text excerpts, $546$ corresponding to the author, and $546$ corresponding to other contemporary authors; and a test set of $272$ excerpts, $136$ belonging to Bront\"{e}, and $136$ written by other authors. The excerpts corresponding to the training and test set, written by either Bront\"{e} or other contemporary authors, are chosen uniformly at random. The objective is to decide if the excerpts in the test set were written by Bront\"{e}.

Again, we consider the three GNN architectures proposed in this paper, as well as the graph coarsening GNN of \cite{defferrard17-cnngraphs}. For the selection GNN, we consider a two-layer architecture, where we choose $100$ nodes (functional words) as determined by each of the three sampling strategies (degree, EDS and SP). For each layer we set $F_{1}=F_{2}=32$, $K_{1}=K_{2}=5$ and $\alpha_{1}=2$ and $\alpha_{2}=4$. In the aggregation GNN we consider three layers, after aggregating all the information at the chosen node (the choice depends on each sampling strategy). In the first layer we compute $F_{1}=32$ features with a filter of size $K_{1}=6$, and do max-pooling, reducing the number of samples by $4$. The second and third layers use filters of size $K_{2}=K_{3}=4$ to obtain $F_{2}=64$ and $F_{3}=128$ features respectively. Pooling is applied, reducing the size of the vector by a factor of $2$ in each of the last two aggregation GNN layers. The multinode GNN employed consists of two outer layers, choosing $P_{1}=30$ and $P_{2}=10$ nodes, respectively, and aggregating information up to the $Q_{1}=Q_{2}=5$ hop-neighborhood. For each outer layer, we have two inner layers, having $16$ features on each of those for the first outer layer, and $16$ and $32$ features for the second outer layer. All filters are of size $3$ and pooling reduces the size of the vectors by half. Finally, the graph coarsening GNN consists of two layers obtaining $F_{1}=F_{2}=32$ features in each, with graph filters of size $K_{1}=K_{2}=5$, and pooling reducing the size of the graph by half on each layer.

\begin{table}
	\centering
\begin{tabular}{lc} \hline
Architecture 				& Accuracy 				\\ \hline
Selection (S) Degree		& $69.6 (\pm 5.6) \%$	\\
Selection (S) EDS			& $68.1 (\pm 5.3) \%$	\\
Selection (S) SP			& $73.0 (\pm 4.8) \%$	\\
Aggregation	(A)	Degree		& $69.5 (\pm 2.0) \%$	\\
Aggregation	(A)	EDS			& $71.0 (\pm 2.8) \%$	\\
Aggregation	(A)	SP			& $69.2 (\pm 4.0) \%$	\\
Multinode (MN) Degree		& $80.4 (\pm 2.0) \%$	\\
\textbf{Multinode (MN) EDS}			& $\mathbf{80.5 (\pm 2.6) \%}$	\\
Multinode (MN) SP			& $79.9 (\pm 2.8) \%$	\\
Graph Coarsening (C) Clustering		& $65.2 (\pm 5.0)\%$ \\ \hline
\end{tabular}
	\caption{Classification accuracy averaged across $10$ different realizations of the training and test sets (recall that the training and test sets are chosen at random from the available corpus, and the choice of training set affects the constructed underlying graph). In parenthesis, we show the standard deviation of the classification accuracy.}
	\label{table_author}
\end{table}

The graph shift operator $\bbS$ is set to the adjacency matrix after normalizing the weights of each row (to add up to $1$) and symmetrizing it, except for the case of graph coarsening GNNs, where the GSO is the normalized Laplacian obtained from the aforementioned adjacency matrix. For training we use $80$ epochs. And we run the experiment $10$ times, to account for the randomness in the selection of training and test sets (and thus, for the randomness in the creation of the underlying WAN).

Results can be found in Table \ref{table_author}, where we show the classification accuracy averaged over $10$ different realizations of the training and test sets, as well as the estimated standard deviation. We first observe that, in this case, all proposed GNN architectures outperform the graph coarsening GNN. We note that the multinode GNN is the best performing architecture. We also observe that selecting nodes via the EDS sampling method works best for aggregation and multinode GNNs, but spectral proxies yield better results in the case of selection GNN. The best classification accuracy obtained is $80.5\%$, on average across all realizations, and achieved by the multinode GNN whose nodes are selected by means of EDS sampling.


\subsection{\texttt{20NEWS} dataset} \label{subsec_20news}

\begin{table}
	\centering
\begin{tabular}{lc} \hline
Architecture 				& Accuracy 				\\ \hline
Selection (S) Degree		& $55.7 (\pm 0.5) \%$	\\
Selection (S) EDS			& $58.1 (\pm 0.5) \%$	\\
Selection (S) SP			& $59.2 (\pm 0.4) \%$	\\
Aggregation	(A)	Degree		& $49.0 (\pm 0.4) \%$	\\
Aggregation	(A)	EDS			& $51.3 (\pm 0.5) \%$	\\
Aggregation	(A)	SP			& $52.9 (\pm 0.5) \%$	\\
Multinode (MN) Degree		& $65.7 (\pm 0.4) \%$	\\
Multinode (MN) EDS			& $66.5 (\pm 0.5) \%$	\\
\textbf{Multinode (MN) SP}			& $\mathbf{67.0 (\pm 0.5) \%}$	\\
Graph Coarsening (C) Clustering		& $62.8 (\pm 0.5)\%$ \\ \hline
\end{tabular}
	\caption{\texttt{20NEWS} dataset on a \texttt{word2vec} graph embedding of $N=1,000$ nodes. Classification accuracy averaged across $10$ different runs. In parenthesis, we show the standard deviation of the classification accuracy.}
	\label{table_20news}
\end{table}

Finally, we consider the classification of articles in the \texttt{20NEWS} dataset which consists of $16,617$ texts ($9,922$ of which are used for training and $6,695$ for testing) \cite{joachims96-20news}. The graph signals are constructed as in \cite{defferrard17-cnngraphs}: each document $x$ is represented using a normalized bag-of-words model and the underlying graph support is constructed using a $16$-NN graph on the \texttt{word2vec} embedding \cite{mikolov13-word2vec} considering the $1,000$ most common words. The GSO adopted is the normalized Laplacian as in \cite{defferrard17-cnngraphs}.

The selection GNN architecture consists of $2$ convolutional layers, selecting $P_{1}=250$ and $P_{2} = 100$ nodes, according to each of the three different sampling strategies. Each layer uses graph filters of $K_{1}=K_{2}=5$ taps to build $F_{1}=32$ and $F_{2}=64$ features. The pooling neighborhoods correspond to $\alpha_{1}=7$ and $\alpha_{2}=12$. For the aggregation GNN we also consider $2$ layers, and use filters of length $K_{1}=K_{2}=11$ to build $F_{1}=F_{2}=32$ features on each layer. Pooling size is $4$, and the data is aggregated at a single node chosen by each of the sampling strategies. The multinode GNN consists of $2$ outer layers that select $P_{1}=70$ and $P_{2}=30$ nodes, respectively. The number of local exchanges to create a temporally-structured signal are $Q_{1} = 10$ and $Q_{2}=25$. Each outer layer employs a regular CNN with $2$ inner layers. Each inner layer of the first outer layer creates $16$ features, while each inner layer of the second outer layer uses $16$ and $32$ features, respectively. All filters involved are of length $5$ and the pooling size is $4$. Finally, for the graph coarsening architecture, we consider $2$ layers, reducing the number of nodes by half on each layer, and computing $F_{1}=32$ and $F_{2}=64$ features, using filters of length $K_{1}=K_{2}=5$.

Training is done for $80$ epochs. Classification accuracy results, averaged out of $10$ runs, are listed in Table~\ref{table_20news}. We note that the multinode GNN is the best performing architecture, followed by graph coarsening. The comparatively poor performance of the aggregation GNN is most likely due to the numerical instabilities that arise from performing a large number of neighborhood exchanges.


\section{Conclusions} \label{sec_conclusions}



In this paper we proposed two architectures for extending convolutional neural networks to process graph signals. The selection graph neural network replaces the convolution operation with graph filtering by means of linear shift invariant graph filters. Pooling is reinterpreted as a neighborhood summarizing function that gathers the relevant regional information at a subset of nodes, followed by a downsampling. By keeping track of the location of these subsets of nodes in the original graph, convolutional layers can be further computed at deeper layers through the use of zero padding. In this way, the selection GNN respects the original topology that describes the data, while reducing the computational complexity at each layer. Furthermore, the resultant features at each layer can be appropriately analyzed in terms of the original graph (frequency analysis, local filtering).

The aggregation GNN collects, at a single node, diffused versions of the original signal. The resulting signal simultaneously possesses a regular temporal structure and includes all relevant information of the topology of the graph. Since the signal collected at this single node has a temporal structure, a regular CNN can be applied to it. In large scale networks, however, gathering all the information of the graph signal at a single node might be infeasible. In order to overcome this, we proposed a multinode variation of the aggregation GNN in which we use a subset of nodes to subsequently create meaningful features of increasing neighborhoods.

We have tested the proposed architectures in a source localization problem on both synthetic and real datasets, as well as for authorship attribution and the classification of articles of the \texttt{20NEWS} dataset. We considered three different ways of choosing nodes in each architecture, based on three existing sampling techniques (namely, by degree, and by leverage scores computed from experimentally designed sampling and spectral proxies). We compared the results with an existing graph coarsening GNN that employs multiscale hierarchical clustering for the pooling stage. We observe that the multinode aggregation GNN exhibits the best performance.

All in all, the proposed GNN architectures exploit the advances in graph signal processing to present novel constructions of deep learning that are able to handle network data represented as signals supported on graphs.

\bibliographystyle{IEEEtran}
\bibliography{myIEEEabrv,biblioArchitGNN}

\end{document}